\documentclass[prl,twocolumn]{revtex4-2}

\usepackage{bm,url,natbib,textcase,color,hyperref,amsmath,amssymb,graphicx,slashed}
\usepackage[utf8]{inputenc}

\begin{document}

{\phantom{.}\vspace{-2.5cm}\\\flushright Imperial-TP-KM-2024-02\\ }

\title{Interactions of massless fermionic fields in three dimensions}

\author{Stefan Fredenhagen}
\email{stefan.fredenhagen@univie.ac.at}

\affiliation{University of Vienna, Faculty of Physics,
Boltzmanngasse 5, 1090 Vienna, Austria}
\affiliation{Erwin Schr\"odinger International Institute for Mathematics and Physics,
University of Vienna, Boltzmanngasse 9, 1090 Vienna, Austria}

\author{Filipp Lausch}
\email{filipp.lausch@univie.ac.at}
\affiliation{University of Vienna, Faculty of Physics,
Boltzmanngasse 5, 1090 Vienna, Austria}

\author{Karapet Mkrtchyan} 
\email{k.mkrtchyan@imperial.ac.uk}
\affiliation{Theoretical Physics Group, Blackett Laboratory, Imperial College London, Prince Consort Road, SW7 2AZ, UK}
\affiliation{Max Planck Institute for Gravitational Physics (Albert Einstein Institute), Am M\"uhlenberg 1, 14476 Potsdam, Germany}

\begin{abstract}
All independent interaction vertices involving massless (Fang--Fronsdal) fermions in three dimensions are classified, completing the classification of interactions of massless fields of any spin. Similarly to the bosonic case, we get no independent vertices at quartic or higher order in the fields involving fields with spin $s\geq 3/2$, and cubic vertices only for spins satisfying triangle inequalities, apart from the cases involving (matter) fields with spin $s\leq 1$. Differently from the bosonic cases, we get only one vertex for each triple of spins with two Majorana fermions and one boson, which is parity even (odd) when the sum of the spins of all fields involved is odd (even). When the two Majorana fermions are identical, their coupling to an odd-spin boson is trivialized. We comment on the non-trivial holographic dictionary relating these vertices to $CFT$ correlators in two dimensions.
\end{abstract}

\maketitle

\section{Introduction}

In this letter, we complete the classification of all independent interactions involving massless (Fang-)Fronsdal fields \cite{Fronsdal:1978rb,Fang:1978wz} of any spin in three-dimensional Einstein spacetimes.
More precisely, we extend the preceding results \cite{Mkrtchyan:2017ixk,Kessel:2018ugi,Fredenhagen:2018guf,Fredenhagen:2019hvb} for interaction vertices in 3D higher-spin (HS) gravity \footnote{Here we use the term ``HS gravity'' for any consistent full non-linear theory that includes (some modification of) Einstein gravity and at least one massless field of spin $s>2$ in its spectrum.}
in metric-like formulation \cite{Campoleoni:2012hp,Fredenhagen:2014oua} to involve fermionic (Fang--Fronsdal) fields \cite{Fang:1978wz}, thus completing the classification of all independent interaction vertices involving bosonic Fronsdal \cite{Fronsdal:1978rb} and fermionic Fang--Fronsdal \cite{Fang:1978wz} massless fields in three dimensions. A similar complete classification for bosonic fields and partial classification for fermionic fields is available for dimensions $d\geq 4$
\footnote{See \cite{Manvelyan:2010jr,Sagnotti:2010at,Joung:2011ww,Henneaux:2012wg,Henneaux:2013gba,Conde:2016izb,Francia:2016weg,Sleight:2017cax,Joung:2019wbl,Fredenhagen:2019lsz} for some key references on Fronsdal program in dimensions $d\geq 4$, and related parallel programs in light-cone approach \cite{Bengtsson:1983pd,Bengtsson:1986kh,Fradkin:1991iy,Metsaev:2005ar,Metsaev:2007rn} and Fradkin-Vasiliev (frame-like) approach \cite{Fradkin:1986qy,Vasiliev:2011knf,Boulanger:2012dx,Khabarov:2020bgr}. See also \cite{Sleight:2016hyl,Ponomarev:2022vjb,Bengtsson:2023ucs,Pekar:2023nev,Campoleoni:2024ced} for recent reviews.}.

To quote Fang and Fronsdal \cite{Fang:1978wz}, ``This paper complements our study of massless
fields with higher, integer spins and demonstrates
that the main results have close analogs in the
case of half-integral spins. The motivation for
our work will not be repeated here.''
We will, however, provide a brief motivation here. 

HS gravities are promising candidates for Quantum Gravity, in particular, as holographic duals (see, e.g., \cite{Klebanov:2002ja,Giombi:2012ms,Gaberdiel:2012uj}) of well-understood CFTs (see, e.g., \cite{Zamolodchikov:1985wn,Bilal:1988jg,Lukyanov:1990tf,Lang:1992zw}). The main problem with known HS gravities \cite{Vasiliev:1990en,Prokushkin:1998bq,Vasiliev:2003ev} is the lack of conventional action principles for these theories \footnote{See, however, \cite{Bonezzi:2015igv} and references therein}.

The analysis of higher-order local completions of non-abelian cubic vertices for higher spins has shown an obstruction at quartic order in $d\geq 4$ (see, e.g.,  \cite{Metsaev:1991mt,Metsaev:1991nb,Dempster:2012vw,Bengtsson:2016hss,Taronna:2017wbx,Roiban:2017iqg,Sleight:2017pcz,Gelfond:2023fwe}).

In three dimensions, the situation is somewhat different. There is a clear distinction between the HS gravities with propagating degrees of freedom \cite{Prokushkin:1998bq,Korybut:2022kdx}, where infinite tower of HS fields interact with matter, but an action principle is not available so far, and those describing only edge modes (see, e.g., \cite{Blencowe:1988gj,Vasiliev:1989re,Campoleoni:2010zq,Henneaux:2010xg,Campoleoni:2011hg,Afshar:2013vka}), which can, in general, be given a Chern--Simons action formulation under mild assumptions \cite{Grigoriev:2020lzu}.
In the case when propagating degrees of freedom are present, the situation is analogous to higher dimensions, with a few key differences, listed below.
\begin{itemize}
    \item Minimal coupling to gravity is available in $d=3$ (as we show here, also for fermionic fields) for massless fields of any spin, thus avoiding the Aragone--Deser problem \cite{Aragone:1979hx,Porrati:2012rd}, as opposed to higher dimensions.
    \item Interactions do not display differences in flat and $(A)dS$ backgrounds (furthermore, they are available in any Einstein background), as opposed to higher dimensions \cite{Fradkin:1987ks,Boulanger:2008tg,Joung:2013nma}. This allows to study HS gravity around flat spacetime \cite{Afshar:2013vka,Gonzalez:2013oaa,Ammon:2020fxs}, which offers a technical simplification in metric-like language.
    \item The existence of a Chern--Simons action indicates that at least for the HS (topological) sector, there may exist also full non-linear (Fang-)Fronsdal Lagrangians for a large class of models, which are yet to be constructed \cite{Neckam:2023}, \footnote{S. Fredenhagen, K.Mkrtchyan, P. Neckam, work in progress.\label{FKN}}.
    \item In the metric-like (Fang-)Fronsdal formulation, it is known how to realize the coupling to matter at least in the lowest order, as opposed to the Chern--Simons setup.
\end{itemize}
All of these properties single out three-dimensional Minkowski spacetime as the simplest playground where the Fronsdal program can be successfully completed, leading to a non-linear theory of HS gravity with matter, with further lessons for higher-dimensional HS gravities.

\section{Preliminaries}

We work with the mostly plus convention for the metric, and we choose a real representation of the gamma-matrices,
\begin{align}
    \gamma^0&=i\,\sigma_2=-\gamma_0\,,\ &\gamma^1&=\sigma_3=\gamma_1\,,\ & \gamma^2=\ &\sigma_1=\gamma_2\,,
\end{align}
where $\sigma$'s are Pauli matrices. They satisfy the gamma-matrix algebra
\begin{align}
    \gamma^\mu \gamma^\nu = \eta^{\mu\nu}+\epsilon^{\mu\nu\rho}\gamma_\rho\,,\label{R1000}
\end{align}
where $\epsilon^{\mu\nu\rho}$ is the Levi--Civita antisymmetric tensor with $\epsilon_{012}=1=-\,\epsilon^{012}$ \footnote{Note that in terms of the generalized Kronecker delta $\delta_{\beta_1\dots\beta_n}^{\alpha_1\dots\alpha_n} = n!\delta^{\alpha_1}_{[ \beta_1}\dots \delta^{\alpha_n}_{\beta_n ]}$ this means $\epsilon_{\mu\nu\rho}=\delta_{\mu\nu\rho}^{012}$ and $\epsilon^{\mu\nu\rho}=-\delta^{\mu\nu\rho}_{012}$.}. Throughout the letter, we use real Majorana spinors $\psi$ satisfying
\begin{equation}
    \psi=\psi_c= C\,(\overline{\psi})^T = C\, (-\psi^\dagger \gamma^0)^T = -C\,(\gamma^0)^T \psi^* =\psi^*\,,
\end{equation}
where the charge conjugation matrix is $C=-\gamma^0$ in this basis. The formulas we use, however, are also valid for Dirac spinors.

We will work in the realm of perturbative field theory, where any interactions can be switched off consistently, by continuously sending corresponding coupling constants to zero. In this case, the independent vertices are those that involve independent coupling constants that can be continuously sent to zero without affecting any lower-order vertex \footnote{In general, sending some of the coupling constants to zero might not be possible in the full non-linear theory, where the independent coupling constants may be related for the consistency of the full theory. However, for the classification of all nontrivial deformations of free action, we do not require the existence of a full nonlinear theory that uses the vertices thus found.}. This way, we can classify all independent $(n+2)-$th order vertices by the lowest-order Noether identity they satisfy (see, e.g., \cite{Fredenhagen:2019lsz}):
\begin{align}
    \delta^{(n)} S^{(2)} + \delta^{(0)} S^{(n+2)}=0\,,\label{Sn}
\end{align}
where the superscripts indicate the order in fields for both the action and gauge transformation:
\begin{align}
    S=S^{(2)}+S^{(3)}+\dots\,,\\
    \delta=\delta^{(0)}+\delta^{(1)}+\dots\,.
\end{align}
We will study the equation \eqref{Sn} for any $n$, or rather its weaker version,
\begin{align}
    \delta^{(0)} S^{(n+2)}\approx 0\,,
\end{align}
where $\approx$ denotes equivalence on free shell, to classify all independent vertices involving fermionic fields in three spacetime dimensions.

The lowest order of the action, $S^{(2)}\,,$ and gauge transformations, $\delta^{(0)}\,,$ are given by (a sum of) (Fang-)Fronsdal actions and corresponding gauge symmetries \cite{Fronsdal:1978rb,Fang:1978wz}:
\begin{align}
    S^{(2)}=\sum_i \int d^3x \,\mathcal{L}^{(2)}(\phi_i)\,,
\end{align}
where the sum goes over all the fields labelled by $i$, with spins $s_i$ taking integer or half-integer values \footnote{With a small abuse of the terminology, we use the term spin $s$ for the massless Fronsdal fields in three dimensions, described by symmetric tensors of rank $s$, and correspondingly, spin $s+1/2$, for Fang--Fronsdal fields, described by symmetric tensor-spinors of rank $s$.}.

For bosonic field $\varphi^{(s)}_{\mu_1\dots\mu_s}$ and fermionic field $\Psi^{(\ell+1/2)}_{\mu_1\dots\mu_\ell}$ with spins $s$ and $\ell+1/2$ respectively, the (Fang-)Fronsdal free Lagrangian $\mathcal{L}^{(2)}$ takes the following form:
\begin{align}
    \mathcal{L}^{(2)}(\varphi^{(s)})=\frac12\,\varphi^{(s)}_{\mu_1\dots \mu_s}\Box \varphi^{(s)\mu_1\dots \mu_s}+\dots\,,\\
    \mathcal{L}^{(2)}(\Psi^{(\ell+1/2)})=\bar{\Psi}^{(\ell+\frac12)}_{\mu_1\dots\mu_\ell}\slashed{\partial}\Psi^{(\ell+\frac12)\mu_1\dots\mu_\ell}+\dots,
\end{align}
where $\slashed{\partial}=\gamma^\mu\partial_\mu$ and the omitted terms contain divergences and (gamma-)traces of the field which are fixed by the condition that the free action is invariant under free gauge transformations \footnote{We use round brackets for symmetrization with weight one (e.g., $A_{(\mu\nu)}=\frac12 (A_{\mu\nu}+A_{\nu\mu})$), and square brackets for anti-symmetrization with weight one (e.g., $A_{[\mu\nu]}=\frac12 (A_{\mu\nu}-A_{\nu\mu})$).}:
\begin{align}
    \delta^{(0)}\varphi^{(s)}_{\mu_1\dots \mu_s}=s\,\partial_{(\mu_1}\epsilon_{\mu_2\dots\mu_s)}\,,\\
    \delta^{(0)}\Psi^{(\ell+\frac12)}_{\mu_1\dots\mu_\ell}=\ell\, \partial_{(\mu_1}\xi_{\mu_2\dots\mu_{\ell})}\,,
\end{align}
where we omit the spinor indices of $\Psi^{(\ell+\frac12)}$ and $\xi$.

As in the bosonic case \cite{Mkrtchyan:2017ixk,Kessel:2018ugi,Fredenhagen:2019hvb}, for classification of the interactions, it is enough to restrict ourselves to the traceless-transverse part of the vertices. Then, it is sufficient to use the free Fierz equations (see details in \cite{Kessel:2018ugi}):
\begin{align}
    \eta^{\mu_1\mu_2}\phi_{\mu_1\mu_2...\mu_s} &= 0, & \partial^{\mu_1}\phi_{\mu_1...\mu_s} &= 0, & \Box\phi_{\mu_1...\mu_s} &= 0,\label{FierzB}
\end{align}
and their fermionic counterparts:
\begin{align}
\gamma^{\mu_1}\Psi_{\mu_1...\mu_\ell} &= 0, & \partial^{\mu_1}\Psi_{\mu_1...\mu_\ell} &= 0, & \slashed{\partial} \Psi_{\mu_1...\mu_\ell} &= 0\,.\label{FierzF1}
\end{align}
As a consequence of \eqref{FierzF1}, we have also:
\begin{align}
    \eta^{\mu_1\mu_2}\Psi_{\mu_1\mu_2...\mu_s} &= 0, & \Box \Psi_{\mu_1...\mu_s} = 0.\label{FierzF2}
\end{align}
When now studying the TT-parts of the vertices for massless fields, we impose gauge symmetry with parameters that also satisfy the Fierz equations. The full off-shell vertices can be derived from the TT part by relaxing these conditions, which will not be pursued here. 
By applying \eqref{R1000} to a fermionic field and using~\eqref{FierzF1}, we can deduce (hereafter $\approx$ is equivalence due to \eqref{FierzB} and \eqref{FierzF1})
\begin{align}
    \Psi^{\mu_1}{}_{\mu_2...\mu_\ell} \approx \epsilon^{\mu_1\alpha\beta}\gamma_\alpha\Psi_{\beta\mu_2...\mu_\ell}, \label{R1}
\end{align}
which will play a major role throughout this paper. Taking the conjugate of the latter equation and using that
\begin{align}
    \gamma_\mu^\dagger=\gamma^0 \gamma_\mu \gamma^0\,,
\end{align}
we find a similar relation for the conjugate field $\Bar{\Psi}_{\mu_1...\mu_s}=-\Psi_{\mu_1...\mu_s}^{\dagger}\gamma^0$,
\begin{align}
    \bar{\Psi}^{\mu_1}{}_{\mu_2...\mu_\ell} = -\epsilon^{\mu_1\alpha\beta}\bar{\Psi}_{\beta\mu_2...\mu_\ell}\gamma_\alpha\,.\label{R1b} 
\end{align}
Another relation can be derived by contracting \eqref{R1000} with a derivative acting on a fermionic field and using Fierz equations \eqref{FierzF1}:
\begin{align}
    0\approx \gamma^\mu \slashed{\partial} \Psi_{\nu_1\dots\nu_\ell} = (\partial^\mu + \epsilon^{\mu\nu\lambda}\gamma_{\lambda}\partial_\nu)\Psi_{\nu_1\dots\nu_\ell}\,,
\end{align}
which further implies:
\begin{align}
    0\approx \epsilon^{\alpha\mu\nu}\partial_{\mu}\Psi_{\nu\nu_2\dots\nu_\ell}+\delta^{\alpha\beta}_{\nu\lambda}\gamma^\lambda\partial^\nu\Psi_{\beta\nu_2\dots\nu_\ell}\nonumber\\
    \approx \epsilon^{\alpha\mu\nu}\partial_{\mu}\Psi_{\nu\nu_2\dots\nu_\ell}\,,
\end{align}
which leads to the following key identities:
\begin{align}
    \partial_{[\mu}\Psi_{\nu]\nu_2\dots\nu_\ell}\approx 0\approx \partial_{[\mu}\bar{\Psi}_{\nu]\nu_2\dots\nu_\ell}\,.\label{CurlEq}
\end{align}

As usual (see, e.g., \cite{Fredenhagen:2019lsz}), when working with symmetric tensor fields, it is useful to introduce an auxiliary vector for each field, to contract the symmetric indices as:
\begin{align}
\phi_i(x_i,a_i) := \frac{1}{s\text{!}}\phi_{\mu_1\dots \mu_{s_i}}(x_i)a_i^{\mu_1}\cdots a_i^{\mu_{s_i}}\,, \\ \Psi_i(x_i,a_i) := \frac{1}{s\text{!}}\Psi_{\mu_1\dots\mu_{\ell_i}}(x_i)a_i^{\mu_1}\cdots a_i^{\mu_{\ell_i}}\label{P1}\,.
\end{align}
Then, one establishes an effective way of contracting Lorentz-indices by introducing operators $A_i^\mu:=\partial_{a_i^\mu}$, which combined with $P_i:=\partial_{x_i^\mu}$ construct the building blocks of vertex operators via scalar contractions:
\begin{align}
s_{ij}&:=P_i\cdot P_j\big|_{1\leq i\leq j\leq n}, & y_{ij}&:=A_i\cdot P_j\big|_{1\leq i,j \leq n},\nonumber \\
z_{ij}&:= A_i\cdot A_j\big|_{1\leq i\leq j\leq n}\,, & \Gamma_{(k)j} & := \gamma_{(k)}^\mu\,A_{j\mu}\,, \label{P2}\\
\Omega_{(k)j} & :=\gamma_{(k)}^{\mu}\,P_{j\mu}\,,\nonumber
\end{align}
where $\gamma_{(k)}$ is a gamma matrix contracting spinor indices of the $k-$th fermion pair in the vertex. We are looking for local scalar vertices of $2f$ fermionic and $n$ bosonic fields given as (the subscript of the Lagrangian term denotes the spins of involved fields)
\begin{align}
&\mathcal{L}_{\ell_1+1/2,...,\ell_{2f}+1/2,s_1,...,s_n} = \mathcal{V}(s_{lm},y_{lm},z_{lm},\Gamma_{(k)l},\Omega_{(k)l})\times\nonumber\\
&\left( \prod_{k=1}^{f}\bar{\Psi}_{2k-1}(x_{2k-1},a_{2k-1})\Psi_{2k}(x_{2k},a_{2k})\right)\times\nonumber\\&\left( \prod_{j=1}^{n}\phi_j(x_{2f+j},a_{2f+j})\right)\bigg|_{\substack{a_i=0\, (i=1,\dots,2f+n)\\ x_i=x\, (i=1,\dots,2f+n)}}\,.\label{P5}
\end{align}
The gauge invariance of the vertices will impose severe restrictions on vertex operators $\mathcal{V}$ as we show below.
Note, that there are also operators involving the tensor $\epsilon^{\mu\nu\rho}$. As we show in the companion paper \cite{FLM2}, these operators can be related to the parity-even operators via relations \eqref{R1} and \eqref{R1b}.

There are several relations to be taken into account. First, the total derivative terms are discarded, which implies the following equations (from now on, we assume the vertex operators to be in the vertex term \eqref{P5}):
\begin{align}
    \sum_{j=1}^{2f+n} s_{ij}\approx 0\,, \quad \sum_{j=1}^{2f+n} y_{ij}\approx 0\,,\quad \sum_{j=1}^{2f+n} \Omega_{(k)j}\approx 0\,.\label{TotalDer}
\end{align}
Next, we impose the Fierz identities \eqref{FierzB}, \eqref{FierzF1} and \eqref{FierzF2} (no summation over repeating indices):
\begin{align}
    s_{ii}\approx 0\,,\qquad y_{ii}\approx 0\,, \qquad z_{ii}\approx 0\,,\qquad\quad\label{FB}\\
    \Gamma_{(k)2k-1}\approx 0\approx \Gamma_{(k)2k}\,,\quad  \Omega_{(k)2k-1}\approx 0\approx \Omega_{(k)2k}\,.\label{FF}
\end{align}
One has also dimension-dependent Schouten identities that impose further relations involving the operators in the vertex (for details see~\cite{Fredenhagen:2019lsz}). In three dimensions, these identities play a crucial role in determining the non-trivial interacting deformations of bosonic free (Fronsdal) theory \cite{Mkrtchyan:2017ixk,Kessel:2018ugi,Fredenhagen:2019hvb}. They should also be taken into account in the interactions of fermions and will be discussed below.

\section{Cubic vertices with Fermions}

The general construction of the interacting theory order-by-order in the power of fields starts from the classification of cubic vertices (see a lightning review in \cite{Kessel:2018ugi}). The vertex operator for the cubic case depends on the following variables (a full basis due to \eqref{TotalDer} -- \eqref{FF}):
\begin{align}
    y_i\equiv y_{i i+1}\,,\quad z_i\equiv z_{i+1 i-1}\,,\quad \Gamma\equiv\Gamma_{(1)3}\,,
\end{align}
where $i=1,2,3$ and $i\pm 3\equiv i$. Note, that $s_{ij}\approx 0 \, (\forall i,j)$, as in the bosonic case, and there is only one new operator ($\Gamma$) compared to the bosonic case (all other $\Gamma$ and $\Omega$ operators vanish on-shell). Furthermore, we omit the index of $\gamma_{(k)}$ (a cubic vertex has at most one fermion pair).

We will need to solve the following equation:
\begin{align}
    \delta^{(0)}S^{(3)}\approx 0\,,
\end{align}
for the case when the cubic interaction contains two fermionic and one bosonic fields. Given that the fermionic fields satisfy also the Fierz equations for the bosonic fields due to \eqref{FierzF1} and \eqref{FierzF2}, the vertex operator (dimension-dependent) identities for bosonic fields \cite{Mkrtchyan:2017ixk,Kessel:2018ugi,Fredenhagen:2019hvb} can be directly used also for fermions. However, fermionic equations \eqref{FierzF1} are stronger than the bosonic ones and imply more identities to be taken into account. More general analysis, involving the parity-odd structures, will be provided in \cite{FLM2}. Here we make a quick derivation of the identities we need. The equations \eqref{R1} and \eqref{R1b} translate into:
\begin{align}
    A_2^\mu\approx \epsilon^{\mu\alpha\beta}\gamma_\alpha A_{2\,\beta}\,,\qquad A_1^\mu\approx -\epsilon^{\mu\alpha\beta} A_{1\,\beta} \gamma_{\alpha}\,,
\end{align}
whereas \eqref{CurlEq} translates into
\begin{align}
    P_{1\,[\mu}A_{1\,\nu]}\approx 0\approx P_{2 [\mu} A_{2\,\nu]}\,.
\end{align}
Together these relations imply:
\begin{align}
    y_1\,z_1\approx y_2\,z_2\approx -y_3\,z_3\,,\quad y_i\,y_j\approx 0\quad (i\neq j)\,,\label{seedid1}
\end{align}
and
\begin{align}
    \Gamma\,y_1\,z_1\approx \Gamma\,y_2\,z_2\approx -\Gamma\,y_3\,z_3\,,\quad \Gamma\,y_i\,y_j\approx 0\; (i\neq j)\,.\label{seedid2}
\end{align}
These equations turn out to be the seeds for all further relations, including Schouten identities.
The gauge transformation for each field is given by the operator $T_i=a_i^\mu P_{i\,\mu}$ of the symmetrized derivative of the parameter, which commutes with $y,z,\Gamma$ as follows:
\begin{align}
    [y_i, T_j]\approx 0\,,\quad [z_i, T_i]=0\,,\quad [\Gamma, T_i]\approx 0\,,\nonumber\\
    [z_{i\pm 1}, T_i]\approx \pm y_{i\mp 1}\,,\qquad\qquad\label{GaugeCommutators}
\end{align}
Therefore, a gauge-invariant vertex has to satisfy
\begin{align}
    \left(y_{i-1}\frac{\partial}{\partial z_{i+1}}-y_{i+1}\frac{\partial}{\partial z_{i-1}}\right){\cal V}(y,z)\approx 0\,,
\end{align}
which is the same as in the bosonic case in any dimensions. The difference, however, is that now the new relations \eqref{seedid1} and \eqref{seedid2} change the space of functions ${\cal V}$ where we solve these equations. 

Using \eqref{seedid1} and \eqref{seedid2}, it is straightforward to show that vertices containing more than one derivative are trivial when the spins satisfy triangle inequalities. 

When the triangle inequalities between spins are not satisfied, and all three massless fields have spins bigger than one, the requirement of gauge invariance is strong enough to kill all candidate cubic vertices. When one requires only one or two gauge symmetries, some candidate vertices may still survive, which is relevant to the study of interactions involving massive fields, which is out of the scope of this work.

For general spins $s_i > 1$, satisfying triangle inequalities, a general solution is given by a linear combination of the following vertex operators:
\begin{align}
    {\cal V}_E(y,z)&=\Gamma\, y_3\, z_1^{n_1}\,z_2^{n_2}\,z_3^{n_3+1}\,,\label{VE}\\
    {\cal V}_O(y,z)&=y_3\, z_1^{n_1}\,z_2^{n_2}\,z_3^{n_3+1}\,,\label{VO}
\end{align}
where ${\cal V}_E \, ({\cal V}_O)$ is parity-even (odd) and $n_i$ are some non-negative integers.
The spins of the fields are given as $s_1=n_2+n_3+3/2\,,\; s_2=n_1+n_3+3/2$ (for both vertices) and $s_3=n_1+n_2+2$ for the vertex \eqref{VE}, while $s_3=n_1+n_2+1$ for the vertex \eqref{VO}.
It is clear, therefore, that the spins of the gauge parameters, $s_i-1$, satisfy triangle inequalities, which turns into equality only for \eqref{VE} when $n_3=0$. This aspect is analogous to the cubic vertices of bosonic fields.

Note that there is only one vertex for each given triple of spins satisfying triangle inequalities, differently from bosonic cases, where we had one parity-even and one parity-odd vertex.

As with the bosonic vertices, the general structure of cubic vertices is different when fields carrying propagating degrees of freedom are involved (we will call those ``matter fields''). Here, these matter fields can take spin values $s=0,1/2,1$. The special vertices are 
\begin{align}
    {\cal V}_{O}^{(1/2,1/2,s)}&=y_3^s\,,& {\cal V}_{E}^{(1/2,1/2,s)}&=\Gamma\, y_3^{s-1}\,,\\
    {\cal V}_{O}^{(1/2,\ell+1/2,0)}&=y_2^{\ell}\,,&
    {\cal V}_{E}^{(1/2,\ell+1/2,1)}&=\Gamma\, y_2^{\ell-1}\,,\\
    {\cal V}_{O}^{(\ell+1/2,1/2,0)}&=y_1^{\ell}\,,&
    {\cal V}_{E}^{(\ell+1/2,1/2,1)}&=\Gamma\, y_1^{\ell-1}\,.
\end{align}
The last option to consider is when the bosonic field is a Chern--Simons vector field instead of Maxwell one. This case is different as the bosonic field equations are stronger:
\begin{align}\label{CSeom}
    P_{3[\mu}A_{3 \nu]}\approx 0\,.
\end{align}
This in turn induces additional identities, including
\begin{align}
    \Gamma y_1\approx 0\approx \Gamma y_2\,,\qquad y_3 z_3\approx 0\,,
\end{align}
which change the classification, allowing only for a zero-derivative vertex.
At the end, we find one vertex with arbitrary equal-spin fermions:
\begin{align}
    {\cal V}_{CS}^{\ell+1/2,\,\ell+1/2,\,1}=\Gamma\,z_3^\ell\,.
\end{align}
This vertex can be derived by covariantizing the derivatives of the Fang--Fronsdal action with $\partial_\mu \rightarrow \partial_\mu + A_\mu$. It requires charged fermions and is therefore trivial for a single Majorana fermion. 

As with the couplings to Fronsdal fields, there is only one fermionic vertex for the coupling to the Chern--Simons field, while there are two bosonic vertices \cite{Kessel:2018ugi}: one parity-even and one parity-odd.

There is an extra parity-odd vertex for coupling of matter fields with spin $1/2$:
\begin{align}
    {\cal V}_{CS, O}^{1/2,\,1/2,\,1}=y_3\,.
\end{align}

This completes the classification of the cubic vertices involving massless fields of any spin.

\section{Higher-order vertices}

By making use of~\eqref{R1} and~\eqref{R1b}, we can achieve a form of the vertex in which every HS fermion appears together with a gamma matrix. The analysis then follows very closely the arguments in \cite{Fredenhagen:2019hvb}. One can first use Schouten identities (potentially after multiplication with a nonvanishing combination of Mandelstam variables $s_{ij}$) to rewrite the vertex such that there are no contractions between field indices (i.e.\  no variables $z_{ij}$) and no contraction between a field index and a gamma matrix index (i.e.\ no variables $\Gamma_{(k)j}$). In the presence of an epsilon tensor, one can use as in \cite{Fredenhagen:2019hvb} Schouten identities to achieve that the epsilon tensor is contracted only to derivative operators, leading to a gauge invariant combination that can be ignored in the remaining analysis.

Similarly, Schouten identities can be used to express all $y_{ij}$ in terms of $y_{i\,i+1}$ and $y_{i\,i+2}$, and analogously all $\Omega_{(k)j}$ by $\Omega_{(k)2k+1}$ (using that $\Omega_{(k)2k-1}\approx \Omega_{(k)2k}\approx 0$). The vertex then depends on the gauge invariant $s_{ij}$ and $\Omega_{(k)2k+1}$, as well as $y_{i\,i+1}$ and $y_{i\,i+2}$. With respect to the latter two variables, gauge invariance of the vertex then requires that it can depend only on the gauge invariant combination
\begin{equation}
    Y_i:= s_{i\,i+2}\,y_{i\,i+1}-s_{i\,i+1}\,y_{i\,i+2}\, .
\end{equation}
As shown in \cite{Fredenhagen:2019hvb}, for the bosonic fields these new variables satisfy
\begin{equation}
    Y_i^2\approx 0\quad (i=2f+1,\dots, 2f+n)\,,
\end{equation}
and therefore a bosonic field can have at most one index (and is at most a spin-1 field). For the fermionic fields, we have instead
\begin{equation}
    Y_i \approx 0\quad (i=1,\dots,2f)\,,
\end{equation}
which follows from~\eqref{CurlEq}. (The same condition also holds for spin-one Chern--Simons field due to~\eqref{CSeom}.) Interactions therefore cannot include fermionic fields with a Lorentz index, and hence they can only be spin-$\frac{1}{2}$ fields. One concludes that there are no independent HS vertices of quartic or higher order.
A more detailed discussion will be provided in \cite{FLM2} and can also be found in \cite{Lausch:2022}.

\section{Discussion}

In this letter, we have classified and explicitly constructed all independent interaction vertices involving massless fermionic fields in three space-time dimensions. The characteristic feature of fermionic vertices is that they all have one derivative. Similarly to the bosonic vertices, cubic vertices with higher-spin fermions exist only for spins satisfying triangle inequalities. Differently from the bosonic case, where for each collection of spins there is one parity-even and one parity-odd vertex, here we find only one vertex for each triple of spins satisfying triangle inequalities, and the parity symmetry of the vertex depends on the spins. 

A posteriori, our results can also be interpreted as a nontrivial check of the holographic dictionary of the AdS$_3$/CFT$_2$ correspondence. For bosonic fields in three dimensions, a single Fronsdal field corresponds on the boundary to a pair of holomorphic and antiholomorphic currents. The two cubic vertices of different parity that exist for bosonic fields satisfying the triangle inequality correspond to combinations of the corresponding conformal three-point functions for holomorphic and antiholomorphic currents on the boundary. For fermionic fields in three spacetime dimensions, a Majorana Fang--Fronsdal field is dual to one Majorana--Weyl fermion on the boundary (see, e.g. \cite{Campoleoni:2017vds}). Its chirality depends on a discrete choice of how to couple the Fang--Fronsdal fields to the AdS background and boundary conditions. This is connected to two inequivalent representations of the Clifford algebra related by $\gamma^{2}\leftrightarrow -\gamma^{2}$ which, in turn, is related to a chirality change from the point of view of the boundary. Fixing the choice for all Fang--Fronsdal fields in the same way (for example to obtain right-moving boundary fermions), the analysis of the vertices carries through and the existence of precisely one vertex for fields satisfying the triangle inequality corresponds to the existence of a unique structure of a three-point function for holomorphic currents satisfying the same inequality. Analogously to the discussion in \cite{Fredenhagen:2018guf}, this match extends to higher-point functions. For completeness, we will present AdS extensions of the vertices presented here in \cite{FLM2}.

This interpretation can also be compared to the holography of
(HS-)supergravity on AdS in a Chern--Simons formulation \cite{Achucarro:1986uwr,Achucarro:1989gm,Witten:1988hc,Banados:1998pi,Henneaux:1999ib,Hyakutake:2012uv,Henneaux:2012ny}. Whereas for bosonic fields the gauge group is of the product form $G\times G$ with the same factors, one can consider \cite{Gunaydin:1986fe} different supergroup extensions of the two factors leading at the boundary to different sets of left-moving and right-moving fermionic charges. 
\footnote{Note, that the holomorphic factorization is specific to massless fields, corresponding to a trivial representation of one of the simple components of the isometry (see, e.g., \cite{Gwak:2015vfb}).}

The general result~\cite{Grigoriev:2020lzu} suggests that any theory of HS fields without propagating degrees of freedom in three dimensions has a frame-like description as a Chern--Simons theory. The results of this letter are consistent with the extension of this statement to theories with fermions. All of the vertices found here can be translated to frame-like vertices and underlie full non-linear theories of the Chern-Simons form. This, in particular, implies that {\it any theory with massless fermionic HS fields without matter can be described as a Chern--Simons gauge theory with a Lie superalgebra.} The first example of such hypergravity was given in \cite{Aragone:1983sz} (see also \cite{Zinoviev:2014sza,Rahman:2019mra}).

Our analysis, on the other hand, treats propagating (matter) fields and HS fields on the same footing and can be the starting point for the construction of a metric-like theory of HS fields coupled to matter.

\section*{Acknowledgements}
We thank Andrea Campoleoni, Euihun Joung, Jan Rosseel, and Eugene Skvortsov for useful discussions. The work of K.M. was supported by the European Union’s Horizon 2020 Research and Innovation Programme under the Marie Sk\l odowska-Curie Grant
No.\ 844265, UKRI and STFC Consolidated Grants ST/T000791/1 and ST/X000575/1, and Alexander von Humboldt renewed research stay grant.

\bibliography{HSrefs}

\begin{thebibliography}{91}%
\makeatletter
\providecommand \@ifxundefined [1]{%
 \@ifx{#1\undefined}
}%
\providecommand \@ifnum [1]{%
 \ifnum #1\expandafter \@firstoftwo
 \else \expandafter \@secondoftwo
 \fi
}%
\providecommand \@ifx [1]{%
 \ifx #1\expandafter \@firstoftwo
 \else \expandafter \@secondoftwo
 \fi
}%
\providecommand \natexlab [1]{#1}%
\providecommand \enquote  [1]{``#1''}%
\providecommand \bibnamefont  [1]{#1}%
\providecommand \bibfnamefont [1]{#1}%
\providecommand \citenamefont [1]{#1}%
\providecommand \href@noop [0]{\@secondoftwo}%
\providecommand \href [0]{\begingroup \@sanitize@url \@href}%
\providecommand \@href[1]{\@@startlink{#1}\@@href}%
\providecommand \@@href[1]{\endgroup#1\@@endlink}%
\providecommand \@sanitize@url [0]{\catcode `\\12\catcode `\$12\catcode
  `\&12\catcode `\#12\catcode `\^12\catcode `\_12\catcode `\%12\relax}%
\providecommand \@@startlink[1]{}%
\providecommand \@@endlink[0]{}%
\providecommand \url  [0]{\begingroup\@sanitize@url \@url }%
\providecommand \@url [1]{\endgroup\@href {#1}{\urlprefix }}%
\providecommand \urlprefix  [0]{URL }%
\providecommand \Eprint [0]{\href }%
\providecommand \doibase [0]{https://doi.org/}%
\providecommand \selectlanguage [0]{\@gobble}%
\providecommand \bibinfo  [0]{\@secondoftwo}%
\providecommand \bibfield  [0]{\@secondoftwo}%
\providecommand \translation [1]{[#1]}%
\providecommand \BibitemOpen [0]{}%
\providecommand \bibitemStop [0]{}%
\providecommand \bibitemNoStop [0]{.\EOS\space}%
\providecommand \EOS [0]{\spacefactor3000\relax}%
\providecommand \BibitemShut  [1]{\csname bibitem#1\endcsname}%
\let\auto@bib@innerbib\@empty
\bibitem [{\citenamefont {Fronsdal}(1978)}]{Fronsdal:1978rb}%
  \BibitemOpen
  \bibfield  {author} {\bibinfo {author} {\bibfnamefont {C.}~\bibnamefont
  {Fronsdal}},\ }\bibfield  {title} {\bibinfo {title} {{Massless Fields with
  Integer Spin}},\ }\href {https://doi.org/10.1103/PhysRevD.18.3624} {\bibfield
   {journal} {\bibinfo  {journal} {Phys. Rev.}\ }\textbf {\bibinfo {volume}
  {D18}},\ \bibinfo {pages} {3624} (\bibinfo {year} {1978})}\BibitemShut
  {NoStop}%
\bibitem [{\citenamefont {Fang}\ and\ \citenamefont
  {Fronsdal}(1978)}]{Fang:1978wz}%
  \BibitemOpen
  \bibfield  {author} {\bibinfo {author} {\bibfnamefont {J.}~\bibnamefont
  {Fang}}\ and\ \bibinfo {author} {\bibfnamefont {C.}~\bibnamefont
  {Fronsdal}},\ }\bibfield  {title} {\bibinfo {title} {{Massless Fields with
  Half Integral Spin}},\ }\href {https://doi.org/10.1103/PhysRevD.18.3630}
  {\bibfield  {journal} {\bibinfo  {journal} {Phys. Rev.}\ }\textbf {\bibinfo
  {volume} {D18}},\ \bibinfo {pages} {3630} (\bibinfo {year}
  {1978})}\BibitemShut {NoStop}%
\bibitem [{\citenamefont {Mkrtchyan}(2018)}]{Mkrtchyan:2017ixk}%
  \BibitemOpen
  \bibfield  {author} {\bibinfo {author} {\bibfnamefont {K.}~\bibnamefont
  {Mkrtchyan}},\ }\bibfield  {title} {\bibinfo {title} {{Cubic interactions of
  massless bosonic fields in three dimensions}},\ }\href
  {https://doi.org/10.1103/PhysRevLett.120.221601} {\bibfield  {journal}
  {\bibinfo  {journal} {Phys. Rev. Lett.}\ }\textbf {\bibinfo {volume} {120}},\
  \bibinfo {pages} {221601} (\bibinfo {year} {2018})},\ \Eprint
  {https://arxiv.org/abs/1712.10003} {arXiv:1712.10003 [hep-th]} \BibitemShut
  {NoStop}%
\bibitem [{\citenamefont {Kessel}\ and\ \citenamefont
  {Mkrtchyan}(2018)}]{Kessel:2018ugi}%
  \BibitemOpen
  \bibfield  {author} {\bibinfo {author} {\bibfnamefont {P.}~\bibnamefont
  {Kessel}}\ and\ \bibinfo {author} {\bibfnamefont {K.}~\bibnamefont
  {Mkrtchyan}},\ }\bibfield  {title} {\bibinfo {title} {{Cubic interactions of
  massless bosonic fields in three dimensions II: Parity-odd and Chern-Simons
  vertices}},\ }\href {https://doi.org/10.1103/PhysRevD.97.106021} {\bibfield
  {journal} {\bibinfo  {journal} {Phys. Rev. D}\ }\textbf {\bibinfo {volume}
  {97}},\ \bibinfo {pages} {106021} (\bibinfo {year} {2018})},\ \Eprint
  {https://arxiv.org/abs/1803.02737} {arXiv:1803.02737 [hep-th]} \BibitemShut
  {NoStop}%
\bibitem [{\citenamefont {Fredenhagen}\ \emph
  {et~al.}(2019{\natexlab{a}})\citenamefont {Fredenhagen}, \citenamefont
  {Kr\"uger},\ and\ \citenamefont {Mkrtchyan}}]{Fredenhagen:2018guf}%
  \BibitemOpen
  \bibfield  {author} {\bibinfo {author} {\bibfnamefont {S.}~\bibnamefont
  {Fredenhagen}}, \bibinfo {author} {\bibfnamefont {O.}~\bibnamefont
  {Kr\"uger}},\ and\ \bibinfo {author} {\bibfnamefont {K.}~\bibnamefont
  {Mkrtchyan}},\ }\bibfield  {title} {\bibinfo {title} {{Constraints for
  Three-Dimensional Higher-Spin Interactions and Conformal Correlators}},\
  }\href {https://doi.org/10.1103/PhysRevD.100.066019} {\bibfield  {journal}
  {\bibinfo  {journal} {Phys. Rev. D}\ }\textbf {\bibinfo {volume} {100}},\
  \bibinfo {pages} {066019} (\bibinfo {year} {2019}{\natexlab{a}})},\ \Eprint
  {https://arxiv.org/abs/1812.10462} {arXiv:1812.10462 [hep-th]} \BibitemShut
  {NoStop}%
\bibitem [{\citenamefont {Fredenhagen}\ \emph
  {et~al.}(2019{\natexlab{b}})\citenamefont {Fredenhagen}, \citenamefont
  {Kr\"uger},\ and\ \citenamefont {Mkrtchyan}}]{Fredenhagen:2019hvb}%
  \BibitemOpen
  \bibfield  {author} {\bibinfo {author} {\bibfnamefont {S.}~\bibnamefont
  {Fredenhagen}}, \bibinfo {author} {\bibfnamefont {O.}~\bibnamefont
  {Kr\"uger}},\ and\ \bibinfo {author} {\bibfnamefont {K.}~\bibnamefont
  {Mkrtchyan}},\ }\bibfield  {title} {\bibinfo {title} {{Vertex-Constraints in
  3D Higher Spin Theories}},\ }\href
  {https://doi.org/10.1103/PhysRevLett.123.131601} {\bibfield  {journal}
  {\bibinfo  {journal} {Phys. Rev. Lett.}\ }\textbf {\bibinfo {volume} {123}},\
  \bibinfo {pages} {131601} (\bibinfo {year} {2019}{\natexlab{b}})},\ \Eprint
  {https://arxiv.org/abs/1905.00093} {arXiv:1905.00093 [hep-th]} \BibitemShut
  {NoStop}%
\bibitem [{Note1()}]{Note1}%
  \BibitemOpen
  \bibinfo {note} {Here we use the term ``HS gravity'' for any consistent full
  non-linear theory that includes (some modification of) Einstein gravity and
  at least one massless field of spin $s>2$ in its spectrum.}\BibitemShut
  {Stop}%
\bibitem [{\citenamefont {Campoleoni}\ \emph {et~al.}(2013)\citenamefont
  {Campoleoni}, \citenamefont {Fredenhagen}, \citenamefont {Pfenninger},\ and\
  \citenamefont {Theisen}}]{Campoleoni:2012hp}%
  \BibitemOpen
  \bibfield  {author} {\bibinfo {author} {\bibfnamefont {A.}~\bibnamefont
  {Campoleoni}}, \bibinfo {author} {\bibfnamefont {S.}~\bibnamefont
  {Fredenhagen}}, \bibinfo {author} {\bibfnamefont {S.}~\bibnamefont
  {Pfenninger}},\ and\ \bibinfo {author} {\bibfnamefont {S.}~\bibnamefont
  {Theisen}},\ }\bibfield  {title} {\bibinfo {title} {{Towards metric-like
  higher-spin gauge theories in three dimensions}},\ }\href
  {https://doi.org/10.1088/1751-8113/46/21/214017} {\bibfield  {journal}
  {\bibinfo  {journal} {J. Phys. A}\ }\textbf {\bibinfo {volume} {46}},\
  \bibinfo {pages} {214017} (\bibinfo {year} {2013})},\ \Eprint
  {https://arxiv.org/abs/1208.1851} {arXiv:1208.1851 [hep-th]} \BibitemShut
  {NoStop}%
\bibitem [{\citenamefont {Fredenhagen}\ and\ \citenamefont
  {Kessel}(2015)}]{Fredenhagen:2014oua}%
  \BibitemOpen
  \bibfield  {author} {\bibinfo {author} {\bibfnamefont {S.}~\bibnamefont
  {Fredenhagen}}\ and\ \bibinfo {author} {\bibfnamefont {P.}~\bibnamefont
  {Kessel}},\ }\bibfield  {title} {\bibinfo {title} {{Metric- and frame-like
  higher-spin gauge theories in three dimensions}},\ }\href
  {https://doi.org/10.1088/1751-8113/48/3/035402} {\bibfield  {journal}
  {\bibinfo  {journal} {J. Phys. A}\ }\textbf {\bibinfo {volume} {48}},\
  \bibinfo {pages} {035402} (\bibinfo {year} {2015})},\ \Eprint
  {https://arxiv.org/abs/1408.2712} {arXiv:1408.2712 [hep-th]} \BibitemShut
  {NoStop}%
\bibitem [{Note2()}]{Note2}%
  \BibitemOpen
  \bibinfo {note} {See \cite
  {Manvelyan:2010jr,Sagnotti:2010at,Joung:2011ww,Henneaux:2012wg,Henneaux:2013gba,Conde:2016izb,Francia:2016weg,Sleight:2017cax,Joung:2019wbl,Fredenhagen:2019lsz}
  for some key references on Fronsdal program in dimensions $d\geq 4$, and
  related parallel programs in light-cone approach \cite
  {Bengtsson:1983pd,Bengtsson:1986kh,Fradkin:1991iy,Metsaev:2005ar,Metsaev:2007rn}
  and Fradkin-Vasiliev (frame-like) approach \cite
  {Fradkin:1986qy,Vasiliev:2011knf,Boulanger:2012dx,Khabarov:2020bgr}. See also
  \cite
  {Sleight:2016hyl,Ponomarev:2022vjb,Bengtsson:2023ucs,Pekar:2023nev,Campoleoni:2024ced}
  for recent reviews.}\BibitemShut {Stop}%
\bibitem [{\citenamefont {Klebanov}\ and\ \citenamefont
  {Polyakov}(2002)}]{Klebanov:2002ja}%
  \BibitemOpen
  \bibfield  {author} {\bibinfo {author} {\bibfnamefont {I.~R.}\ \bibnamefont
  {Klebanov}}\ and\ \bibinfo {author} {\bibfnamefont {A.~M.}\ \bibnamefont
  {Polyakov}},\ }\bibfield  {title} {\bibinfo {title} {{AdS dual of the
  critical O(N) vector model}},\ }\href
  {https://doi.org/10.1016/S0370-2693(02)02980-5} {\bibfield  {journal}
  {\bibinfo  {journal} {Phys. Lett. B}\ }\textbf {\bibinfo {volume} {550}},\
  \bibinfo {pages} {213} (\bibinfo {year} {2002})},\ \Eprint
  {https://arxiv.org/abs/hep-th/0210114} {arXiv:hep-th/0210114} \BibitemShut
  {NoStop}%
\bibitem [{\citenamefont {Giombi}\ and\ \citenamefont
  {Yin}(2013)}]{Giombi:2012ms}%
  \BibitemOpen
  \bibfield  {author} {\bibinfo {author} {\bibfnamefont {S.}~\bibnamefont
  {Giombi}}\ and\ \bibinfo {author} {\bibfnamefont {X.}~\bibnamefont {Yin}},\
  }\bibfield  {title} {\bibinfo {title} {{The Higher Spin/Vector Model
  Duality}},\ }\href {https://doi.org/10.1088/1751-8113/46/21/214003}
  {\bibfield  {journal} {\bibinfo  {journal} {J. Phys. A}\ }\textbf {\bibinfo
  {volume} {46}},\ \bibinfo {pages} {214003} (\bibinfo {year} {2013})},\
  \Eprint {https://arxiv.org/abs/1208.4036} {arXiv:1208.4036 [hep-th]}
  \BibitemShut {NoStop}%
\bibitem [{\citenamefont {Gaberdiel}\ and\ \citenamefont
  {Gopakumar}(2013)}]{Gaberdiel:2012uj}%
  \BibitemOpen
  \bibfield  {author} {\bibinfo {author} {\bibfnamefont {M.~R.}\ \bibnamefont
  {Gaberdiel}}\ and\ \bibinfo {author} {\bibfnamefont {R.}~\bibnamefont
  {Gopakumar}},\ }\bibfield  {title} {\bibinfo {title} {{Minimal Model
  Holography}},\ }\href {https://doi.org/10.1088/1751-8113/46/21/214002}
  {\bibfield  {journal} {\bibinfo  {journal} {J. Phys. A}\ }\textbf {\bibinfo
  {volume} {46}},\ \bibinfo {pages} {214002} (\bibinfo {year} {2013})},\
  \Eprint {https://arxiv.org/abs/1207.6697} {arXiv:1207.6697 [hep-th]}
  \BibitemShut {NoStop}%
\bibitem [{\citenamefont {Zamolodchikov}(1985)}]{Zamolodchikov:1985wn}%
  \BibitemOpen
  \bibfield  {author} {\bibinfo {author} {\bibfnamefont {A.~B.}\ \bibnamefont
  {Zamolodchikov}},\ }\bibfield  {title} {\bibinfo {title} {{Infinite
  Additional Symmetries in Two-Dimensional Conformal Quantum Field Theory}},\
  }\href {https://doi.org/10.1007/BF01036128} {\bibfield  {journal} {\bibinfo
  {journal} {Theor. Math. Phys.}\ }\textbf {\bibinfo {volume} {65}},\ \bibinfo
  {pages} {1205} (\bibinfo {year} {1985})}\BibitemShut {NoStop}%
\bibitem [{\citenamefont {Bilal}\ and\ \citenamefont
  {Gervais}(1989)}]{Bilal:1988jg}%
  \BibitemOpen
  \bibfield  {author} {\bibinfo {author} {\bibfnamefont {A.}~\bibnamefont
  {Bilal}}\ and\ \bibinfo {author} {\bibfnamefont {J.-L.}\ \bibnamefont
  {Gervais}},\ }\bibfield  {title} {\bibinfo {title} {{Systematic Construction
  of Conformal Theories with Higher Spin Virasoro Symmetries}},\ }\href
  {https://doi.org/10.1016/0550-3213(89)90633-0} {\bibfield  {journal}
  {\bibinfo  {journal} {Nucl. Phys. B}\ }\textbf {\bibinfo {volume} {318}},\
  \bibinfo {pages} {579} (\bibinfo {year} {1989})}\BibitemShut {NoStop}%
\bibitem [{\citenamefont {Lukyanov}\ and\ \citenamefont
  {Fateev}(1990)}]{Lukyanov:1990tf}%
  \BibitemOpen
  \bibfield  {author} {\bibinfo {author} {\bibfnamefont {S.~L.}\ \bibnamefont
  {Lukyanov}}\ and\ \bibinfo {author} {\bibfnamefont {V.~A.}\ \bibnamefont
  {Fateev}},\ }\bibfield  {title} {\bibinfo {title} {{Additional symmetries and
  exactly soluble models in two-dimensional conformal field theory}},\
  }\href@noop {} {\bibfield  {journal} {\bibinfo  {journal} {Soviet Scientific
  Reviews A, Physics: 15.2}\ } (\bibinfo {year} {1990})}\BibitemShut {NoStop}%
\bibitem [{\citenamefont {Lang}\ and\ \citenamefont
  {R{\"u}hl}(1993)}]{Lang:1992zw}%
  \BibitemOpen
  \bibfield  {author} {\bibinfo {author} {\bibfnamefont {K.}~\bibnamefont
  {Lang}}\ and\ \bibinfo {author} {\bibfnamefont {W.}~\bibnamefont
  {R{\"u}hl}},\ }\bibfield  {title} {\bibinfo {title} {{The Critical O(N) sigma
  model at dimensions 2 \ensuremath{<} d \ensuremath{<} 4: Fusion coefficients
  and anomalous dimensions}},\ }\href
  {https://doi.org/10.1016/0550-3213(93)90417-N} {\bibfield  {journal}
  {\bibinfo  {journal} {Nucl. Phys. B}\ }\textbf {\bibinfo {volume} {400}},\
  \bibinfo {pages} {597} (\bibinfo {year} {1993})}\BibitemShut {NoStop}%
\bibitem [{\citenamefont {Vasiliev}(1990)}]{Vasiliev:1990en}%
  \BibitemOpen
  \bibfield  {author} {\bibinfo {author} {\bibfnamefont {M.~A.}\ \bibnamefont
  {Vasiliev}},\ }\bibfield  {title} {\bibinfo {title} {{Consistent equation for
  interacting gauge fields of all spins in (3+1)-dimensions}},\ }\href
  {https://doi.org/10.1016/0370-2693(90)91400-6} {\bibfield  {journal}
  {\bibinfo  {journal} {Phys. Lett.}\ }\textbf {\bibinfo {volume} {B243}},\
  \bibinfo {pages} {378} (\bibinfo {year} {1990})}\BibitemShut {NoStop}%
\bibitem [{\citenamefont {Prokushkin}\ and\ \citenamefont
  {Vasiliev}(1999)}]{Prokushkin:1998bq}%
  \BibitemOpen
  \bibfield  {author} {\bibinfo {author} {\bibfnamefont {S.~F.}\ \bibnamefont
  {Prokushkin}}\ and\ \bibinfo {author} {\bibfnamefont {M.~A.}\ \bibnamefont
  {Vasiliev}},\ }\bibfield  {title} {\bibinfo {title} {{Higher spin gauge
  interactions for massive matter fields in 3-D AdS space-time}},\ }\href
  {https://doi.org/10.1016/S0550-3213(98)00839-6} {\bibfield  {journal}
  {\bibinfo  {journal} {Nucl. Phys. B}\ }\textbf {\bibinfo {volume} {545}},\
  \bibinfo {pages} {385} (\bibinfo {year} {1999})},\ \Eprint
  {https://arxiv.org/abs/hep-th/9806236} {arXiv:hep-th/9806236} \BibitemShut
  {NoStop}%
\bibitem [{\citenamefont {Vasiliev}(2003)}]{Vasiliev:2003ev}%
  \BibitemOpen
  \bibfield  {author} {\bibinfo {author} {\bibfnamefont {M.~A.}\ \bibnamefont
  {Vasiliev}},\ }\bibfield  {title} {\bibinfo {title} {{Nonlinear equations for
  symmetric massless higher spin fields in (A)dS(d)}},\ }\href
  {https://doi.org/10.1016/S0370-2693(03)00872-4} {\bibfield  {journal}
  {\bibinfo  {journal} {Phys. Lett.}\ }\textbf {\bibinfo {volume} {B567}},\
  \bibinfo {pages} {139} (\bibinfo {year} {2003})},\ \Eprint
  {https://arxiv.org/abs/hep-th/0304049} {arXiv:hep-th/0304049 [hep-th]}
  \BibitemShut {NoStop}%
\bibitem [{Note3()}]{Note3}%
  \BibitemOpen
  \bibinfo {note} {See, however, \cite {Bonezzi:2015igv} and references
  therein}\BibitemShut {NoStop}%
\bibitem [{\citenamefont {Metsaev}(1991{\natexlab{a}})}]{Metsaev:1991mt}%
  \BibitemOpen
  \bibfield  {author} {\bibinfo {author} {\bibfnamefont {R.~R.}\ \bibnamefont
  {Metsaev}},\ }\bibfield  {title} {\bibinfo {title} {{Poincare invariant
  dynamics of massless higher spins: Fourth order analysis on mass shell}},\
  }\href {https://doi.org/10.1142/S0217732391000348} {\bibfield  {journal}
  {\bibinfo  {journal} {Mod. Phys. Lett.}\ }\textbf {\bibinfo {volume} {A6}},\
  \bibinfo {pages} {359} (\bibinfo {year} {1991}{\natexlab{a}})}\BibitemShut
  {NoStop}%
\bibitem [{\citenamefont {Metsaev}(1991{\natexlab{b}})}]{Metsaev:1991nb}%
  \BibitemOpen
  \bibfield  {author} {\bibinfo {author} {\bibfnamefont {R.~R.}\ \bibnamefont
  {Metsaev}},\ }\bibfield  {title} {\bibinfo {title} {{S matrix approach to
  massless higher spins theory. 2: The Case of internal symmetry}},\ }\href
  {https://doi.org/10.1142/S0217732391002839} {\bibfield  {journal} {\bibinfo
  {journal} {Mod. Phys. Lett.}\ }\textbf {\bibinfo {volume} {A6}},\ \bibinfo
  {pages} {2411} (\bibinfo {year} {1991}{\natexlab{b}})}\BibitemShut {NoStop}%
\bibitem [{\citenamefont {Dempster}\ and\ \citenamefont
  {Tsulaia}(2012)}]{Dempster:2012vw}%
  \BibitemOpen
  \bibfield  {author} {\bibinfo {author} {\bibfnamefont {P.}~\bibnamefont
  {Dempster}}\ and\ \bibinfo {author} {\bibfnamefont {M.}~\bibnamefont
  {Tsulaia}},\ }\bibfield  {title} {\bibinfo {title} {{On the Structure of
  Quartic Vertices for Massless Higher Spin Fields on Minkowski Background}},\
  }\href {https://doi.org/10.1016/j.nuclphysb.2012.07.031} {\bibfield
  {journal} {\bibinfo  {journal} {Nucl. Phys. B}\ }\textbf {\bibinfo {volume}
  {865}},\ \bibinfo {pages} {353} (\bibinfo {year} {2012})},\ \Eprint
  {https://arxiv.org/abs/1203.5597} {arXiv:1203.5597 [hep-th]} \BibitemShut
  {NoStop}%
\bibitem [{\citenamefont {Bengtsson}(2016)}]{Bengtsson:2016hss}%
  \BibitemOpen
  \bibfield  {author} {\bibinfo {author} {\bibfnamefont {A.~K.~H.}\
  \bibnamefont {Bengtsson}},\ }\bibfield  {title} {\bibinfo {title}
  {{Investigations into Light-front Quartic Interactions for Massless Fields
  (I): Non-constructibility of Higher Spin Quartic Amplitudes}},\ }\href
  {https://doi.org/10.1007/JHEP12(2016)134} {\bibfield  {journal} {\bibinfo
  {journal} {JHEP}\ }\textbf {\bibinfo {volume} {12}},\ \bibinfo {pages}
  {134}},\ \Eprint {https://arxiv.org/abs/1607.06659} {arXiv:1607.06659
  [hep-th]} \BibitemShut {NoStop}%
\bibitem [{\citenamefont {Taronna}(2017)}]{Taronna:2017wbx}%
  \BibitemOpen
  \bibfield  {author} {\bibinfo {author} {\bibfnamefont {M.}~\bibnamefont
  {Taronna}},\ }\bibfield  {title} {\bibinfo {title} {{On the Non-Local
  Obstruction to Interacting Higher Spins in Flat Space}},\ }\href
  {https://doi.org/10.1007/JHEP05(2017)026} {\bibfield  {journal} {\bibinfo
  {journal} {JHEP}\ }\textbf {\bibinfo {volume} {05}},\ \bibinfo {pages}
  {026}},\ \Eprint {https://arxiv.org/abs/1701.05772} {arXiv:1701.05772
  [hep-th]} \BibitemShut {NoStop}%
\bibitem [{\citenamefont {Roiban}\ and\ \citenamefont
  {Tseytlin}(2017)}]{Roiban:2017iqg}%
  \BibitemOpen
  \bibfield  {author} {\bibinfo {author} {\bibfnamefont {R.}~\bibnamefont
  {Roiban}}\ and\ \bibinfo {author} {\bibfnamefont {A.~A.}\ \bibnamefont
  {Tseytlin}},\ }\bibfield  {title} {\bibinfo {title} {{On four-point
  interactions in massless higher spin theory in flat space}},\ }\href
  {https://doi.org/10.1007/JHEP04(2017)139} {\bibfield  {journal} {\bibinfo
  {journal} {JHEP}\ }\textbf {\bibinfo {volume} {04}},\ \bibinfo {pages}
  {139}},\ \Eprint {https://arxiv.org/abs/1701.05773} {arXiv:1701.05773
  [hep-th]} \BibitemShut {NoStop}%
\bibitem [{\citenamefont {Sleight}\ and\ \citenamefont
  {Taronna}(2018{\natexlab{a}})}]{Sleight:2017pcz}%
  \BibitemOpen
  \bibfield  {author} {\bibinfo {author} {\bibfnamefont {C.}~\bibnamefont
  {Sleight}}\ and\ \bibinfo {author} {\bibfnamefont {M.}~\bibnamefont
  {Taronna}},\ }\bibfield  {title} {\bibinfo {title} {{Higher-Spin Gauge
  Theories and Bulk Locality}},\ }\href
  {https://doi.org/10.1103/PhysRevLett.121.171604} {\bibfield  {journal}
  {\bibinfo  {journal} {Phys. Rev. Lett.}\ }\textbf {\bibinfo {volume} {121}},\
  \bibinfo {pages} {171604} (\bibinfo {year} {2018}{\natexlab{a}})},\ \Eprint
  {https://arxiv.org/abs/1704.07859} {arXiv:1704.07859 [hep-th]} \BibitemShut
  {NoStop}%
\bibitem [{\citenamefont {Gelfond}(2023)}]{Gelfond:2023fwe}%
  \BibitemOpen
  \bibfield  {author} {\bibinfo {author} {\bibfnamefont {O.~A.}\ \bibnamefont
  {Gelfond}},\ }\bibfield  {title} {\bibinfo {title} {{Moderately non-local
  $\eta {\bar{\eta }}$ vertices in the $AdS_4$ higher-spin gauge theory}},\
  }\href {https://doi.org/10.1140/epjc/s10052-023-12308-x} {\bibfield
  {journal} {\bibinfo  {journal} {Eur. Phys. J. C}\ }\textbf {\bibinfo {volume}
  {83}},\ \bibinfo {pages} {1154} (\bibinfo {year} {2023})},\ \Eprint
  {https://arxiv.org/abs/2308.16281} {arXiv:2308.16281 [hep-th]} \BibitemShut
  {NoStop}%
\bibitem [{\citenamefont {Korybut}\ \emph {et~al.}(2023)\citenamefont
  {Korybut}, \citenamefont {Sevostyanova}, \citenamefont {Vasiliev},\ and\
  \citenamefont {Vereitin}}]{Korybut:2022kdx}%
  \BibitemOpen
  \bibfield  {author} {\bibinfo {author} {\bibfnamefont {A.~V.}\ \bibnamefont
  {Korybut}}, \bibinfo {author} {\bibfnamefont {A.~A.}\ \bibnamefont
  {Sevostyanova}}, \bibinfo {author} {\bibfnamefont {M.~A.}\ \bibnamefont
  {Vasiliev}},\ and\ \bibinfo {author} {\bibfnamefont {V.~A.}\ \bibnamefont
  {Vereitin}},\ }\bibfield  {title} {\bibinfo {title} {{Disentanglement of
  topological and dynamical fields in 3d higher-spin theory within shifted
  homotopy approach}},\ }\href {https://doi.org/10.1016/j.physletb.2023.137718}
  {\bibfield  {journal} {\bibinfo  {journal} {Phys. Lett. B}\ }\textbf
  {\bibinfo {volume} {838}},\ \bibinfo {pages} {137718} (\bibinfo {year}
  {2023})},\ \Eprint {https://arxiv.org/abs/2211.15778} {arXiv:2211.15778
  [hep-th]} \BibitemShut {NoStop}%
\bibitem [{\citenamefont {Blencowe}(1989)}]{Blencowe:1988gj}%
  \BibitemOpen
  \bibfield  {author} {\bibinfo {author} {\bibfnamefont {M.~P.}\ \bibnamefont
  {Blencowe}},\ }\bibfield  {title} {\bibinfo {title} {{A Consistent
  Interacting Massless Higher Spin Field Theory in $D$ = (2+1)}},\ }\href
  {https://doi.org/10.1088/0264-9381/6/4/005} {\bibfield  {journal} {\bibinfo
  {journal} {Class. Quant. Grav.}\ }\textbf {\bibinfo {volume} {6}},\ \bibinfo
  {pages} {443} (\bibinfo {year} {1989})}\BibitemShut {NoStop}%
\bibitem [{\citenamefont {Vasiliev}(1991)}]{Vasiliev:1989re}%
  \BibitemOpen
  \bibfield  {author} {\bibinfo {author} {\bibfnamefont {M.~A.}\ \bibnamefont
  {Vasiliev}},\ }\bibfield  {title} {\bibinfo {title} {{Higher Spin Algebras
  and Quantization on the Sphere and Hyperboloid}},\ }\href
  {https://doi.org/10.1142/S0217751X91000605} {\bibfield  {journal} {\bibinfo
  {journal} {Int. J. Mod. Phys. A}\ }\textbf {\bibinfo {volume} {6}},\ \bibinfo
  {pages} {1115} (\bibinfo {year} {1991})}\BibitemShut {NoStop}%
\bibitem [{\citenamefont {Campoleoni}\ \emph {et~al.}(2010)\citenamefont
  {Campoleoni}, \citenamefont {Fredenhagen}, \citenamefont {Pfenninger},\ and\
  \citenamefont {Theisen}}]{Campoleoni:2010zq}%
  \BibitemOpen
  \bibfield  {author} {\bibinfo {author} {\bibfnamefont {A.}~\bibnamefont
  {Campoleoni}}, \bibinfo {author} {\bibfnamefont {S.}~\bibnamefont
  {Fredenhagen}}, \bibinfo {author} {\bibfnamefont {S.}~\bibnamefont
  {Pfenninger}},\ and\ \bibinfo {author} {\bibfnamefont {S.}~\bibnamefont
  {Theisen}},\ }\bibfield  {title} {\bibinfo {title} {{Asymptotic symmetries of
  three-dimensional gravity coupled to higher-spin fields}},\ }\href
  {https://doi.org/10.1007/JHEP11(2010)007} {\bibfield  {journal} {\bibinfo
  {journal} {JHEP}\ }\textbf {\bibinfo {volume} {11}},\ \bibinfo {pages}
  {007}},\ \Eprint {https://arxiv.org/abs/1008.4744} {arXiv:1008.4744 [hep-th]}
  \BibitemShut {NoStop}%
\bibitem [{\citenamefont {Henneaux}\ and\ \citenamefont
  {Rey}(2010)}]{Henneaux:2010xg}%
  \BibitemOpen
  \bibfield  {author} {\bibinfo {author} {\bibfnamefont {M.}~\bibnamefont
  {Henneaux}}\ and\ \bibinfo {author} {\bibfnamefont {S.-J.}\ \bibnamefont
  {Rey}},\ }\bibfield  {title} {\bibinfo {title} {{Nonlinear $W_{infinity}$ as
  Asymptotic Symmetry of Three-Dimensional Higher Spin Anti-de Sitter
  Gravity}},\ }\href {https://doi.org/10.1007/JHEP12(2010)007} {\bibfield
  {journal} {\bibinfo  {journal} {JHEP}\ }\textbf {\bibinfo {volume} {12}},\
  \bibinfo {pages} {007}},\ \Eprint {https://arxiv.org/abs/1008.4579}
  {arXiv:1008.4579 [hep-th]} \BibitemShut {NoStop}%
\bibitem [{\citenamefont {Campoleoni}\ \emph {et~al.}(2011)\citenamefont
  {Campoleoni}, \citenamefont {Fredenhagen},\ and\ \citenamefont
  {Pfenninger}}]{Campoleoni:2011hg}%
  \BibitemOpen
  \bibfield  {author} {\bibinfo {author} {\bibfnamefont {A.}~\bibnamefont
  {Campoleoni}}, \bibinfo {author} {\bibfnamefont {S.}~\bibnamefont
  {Fredenhagen}},\ and\ \bibinfo {author} {\bibfnamefont {S.}~\bibnamefont
  {Pfenninger}},\ }\bibfield  {title} {\bibinfo {title} {{Asymptotic
  W-symmetries in three-dimensional higher-spin gauge theories}},\ }\href
  {https://doi.org/10.1007/JHEP09(2011)113} {\bibfield  {journal} {\bibinfo
  {journal} {JHEP}\ }\textbf {\bibinfo {volume} {09}},\ \bibinfo {pages}
  {113}},\ \Eprint {https://arxiv.org/abs/1107.0290} {arXiv:1107.0290 [hep-th]}
  \BibitemShut {NoStop}%
\bibitem [{\citenamefont {Afshar}\ \emph {et~al.}(2013)\citenamefont {Afshar},
  \citenamefont {Bagchi}, \citenamefont {Fareghbal}, \citenamefont
  {Grumiller},\ and\ \citenamefont {Rosseel}}]{Afshar:2013vka}%
  \BibitemOpen
  \bibfield  {author} {\bibinfo {author} {\bibfnamefont {H.}~\bibnamefont
  {Afshar}}, \bibinfo {author} {\bibfnamefont {A.}~\bibnamefont {Bagchi}},
  \bibinfo {author} {\bibfnamefont {R.}~\bibnamefont {Fareghbal}}, \bibinfo
  {author} {\bibfnamefont {D.}~\bibnamefont {Grumiller}},\ and\ \bibinfo
  {author} {\bibfnamefont {J.}~\bibnamefont {Rosseel}},\ }\bibfield  {title}
  {\bibinfo {title} {{Spin-3 Gravity in Three-Dimensional Flat Space}},\ }\href
  {https://doi.org/10.1103/PhysRevLett.111.121603} {\bibfield  {journal}
  {\bibinfo  {journal} {Phys. Rev. Lett.}\ }\textbf {\bibinfo {volume} {111}},\
  \bibinfo {pages} {121603} (\bibinfo {year} {2013})},\ \Eprint
  {https://arxiv.org/abs/1307.4768} {arXiv:1307.4768 [hep-th]} \BibitemShut
  {NoStop}%
\bibitem [{\citenamefont {Grigoriev}\ \emph {et~al.}(2020)\citenamefont
  {Grigoriev}, \citenamefont {Mkrtchyan},\ and\ \citenamefont
  {Skvortsov}}]{Grigoriev:2020lzu}%
  \BibitemOpen
  \bibfield  {author} {\bibinfo {author} {\bibfnamefont {M.}~\bibnamefont
  {Grigoriev}}, \bibinfo {author} {\bibfnamefont {K.}~\bibnamefont
  {Mkrtchyan}},\ and\ \bibinfo {author} {\bibfnamefont {E.}~\bibnamefont
  {Skvortsov}},\ }\bibfield  {title} {\bibinfo {title} {{Matter-free higher
  spin gravities in 3D: Partially-massless fields and general structure}},\
  }\href {https://doi.org/10.1103/PhysRevD.102.066003} {\bibfield  {journal}
  {\bibinfo  {journal} {Phys. Rev. D}\ }\textbf {\bibinfo {volume} {102}},\
  \bibinfo {pages} {066003} (\bibinfo {year} {2020})},\ \Eprint
  {https://arxiv.org/abs/2005.05931} {arXiv:2005.05931 [hep-th]} \BibitemShut
  {NoStop}%
\bibitem [{\citenamefont {Aragone}\ and\ \citenamefont
  {Deser}(1979)}]{Aragone:1979hx}%
  \BibitemOpen
  \bibfield  {author} {\bibinfo {author} {\bibfnamefont {C.}~\bibnamefont
  {Aragone}}\ and\ \bibinfo {author} {\bibfnamefont {S.}~\bibnamefont
  {Deser}},\ }\bibfield  {title} {\bibinfo {title} {{Consistency Problems of
  Hypergravity}},\ }\href {https://doi.org/10.1016/0370-2693(79)90808-6}
  {\bibfield  {journal} {\bibinfo  {journal} {Phys. Lett.}\ }\textbf {\bibinfo
  {volume} {B86}},\ \bibinfo {pages} {161} (\bibinfo {year}
  {1979})}\BibitemShut {NoStop}%
\bibitem [{\citenamefont {Porrati}(2012)}]{Porrati:2012rd}%
  \BibitemOpen
  \bibfield  {author} {\bibinfo {author} {\bibfnamefont {M.}~\bibnamefont
  {Porrati}},\ }\bibfield  {title} {\bibinfo {title} {{Old and New No Go
  Theorems on Interacting Massless Particles in Flat Space}},\ }in\ \href
  {http://inspirehep.net/record/1187634/files/arXiv:1209.4876.pdf} {\emph
  {\bibinfo {booktitle} {{17th International Seminar on High Energy Physics
  (Quarks 2012) Yaroslavl, Russia, June 4-10, 2012}}}}\ (\bibinfo {year}
  {2012})\ \Eprint {https://arxiv.org/abs/1209.4876} {arXiv:1209.4876 [hep-th]}
  \BibitemShut {NoStop}%
\bibitem [{\citenamefont {Fradkin}\ and\ \citenamefont
  {Vasiliev}(1987{\natexlab{a}})}]{Fradkin:1987ks}%
  \BibitemOpen
  \bibfield  {author} {\bibinfo {author} {\bibfnamefont {E.~S.}\ \bibnamefont
  {Fradkin}}\ and\ \bibinfo {author} {\bibfnamefont {M.~A.}\ \bibnamefont
  {Vasiliev}},\ }\bibfield  {title} {\bibinfo {title} {{On the Gravitational
  Interaction of Massless Higher Spin Fields}},\ }\href
  {https://doi.org/10.1016/0370-2693(87)91275-5} {\bibfield  {journal}
  {\bibinfo  {journal} {Phys. Lett.}\ }\textbf {\bibinfo {volume} {B189}},\
  \bibinfo {pages} {89} (\bibinfo {year} {1987}{\natexlab{a}})}\BibitemShut
  {NoStop}%
\bibitem [{\citenamefont {Boulanger}\ \emph {et~al.}(2008)\citenamefont
  {Boulanger}, \citenamefont {Leclercq},\ and\ \citenamefont
  {Sundell}}]{Boulanger:2008tg}%
  \BibitemOpen
  \bibfield  {author} {\bibinfo {author} {\bibfnamefont {N.}~\bibnamefont
  {Boulanger}}, \bibinfo {author} {\bibfnamefont {S.}~\bibnamefont
  {Leclercq}},\ and\ \bibinfo {author} {\bibfnamefont {P.}~\bibnamefont
  {Sundell}},\ }\bibfield  {title} {\bibinfo {title} {{On The Uniqueness of
  Minimal Coupling in Higher-Spin Gauge Theory}},\ }\href
  {https://doi.org/10.1088/1126-6708/2008/08/056} {\bibfield  {journal}
  {\bibinfo  {journal} {JHEP}\ }\textbf {\bibinfo {volume} {08}},\ \bibinfo
  {pages} {056}},\ \Eprint {https://arxiv.org/abs/0805.2764} {arXiv:0805.2764
  [hep-th]} \BibitemShut {NoStop}%
\bibitem [{\citenamefont {Joung}\ and\ \citenamefont
  {Taronna}(2014)}]{Joung:2013nma}%
  \BibitemOpen
  \bibfield  {author} {\bibinfo {author} {\bibfnamefont {E.}~\bibnamefont
  {Joung}}\ and\ \bibinfo {author} {\bibfnamefont {M.}~\bibnamefont
  {Taronna}},\ }\bibfield  {title} {\bibinfo {title}
  {{Cubic-interaction-induced deformations of higher-spin symmetries}},\ }\href
  {https://doi.org/10.1007/JHEP03(2014)103} {\bibfield  {journal} {\bibinfo
  {journal} {JHEP}\ }\textbf {\bibinfo {volume} {03}},\ \bibinfo {pages}
  {103}},\ \Eprint {https://arxiv.org/abs/1311.0242} {arXiv:1311.0242 [hep-th]}
  \BibitemShut {NoStop}%
\bibitem [{\citenamefont {Gonzalez}\ \emph {et~al.}(2013)\citenamefont
  {Gonzalez}, \citenamefont {Matulich}, \citenamefont {Pino},\ and\
  \citenamefont {Troncoso}}]{Gonzalez:2013oaa}%
  \BibitemOpen
  \bibfield  {author} {\bibinfo {author} {\bibfnamefont {H.~A.}\ \bibnamefont
  {Gonzalez}}, \bibinfo {author} {\bibfnamefont {J.}~\bibnamefont {Matulich}},
  \bibinfo {author} {\bibfnamefont {M.}~\bibnamefont {Pino}},\ and\ \bibinfo
  {author} {\bibfnamefont {R.}~\bibnamefont {Troncoso}},\ }\bibfield  {title}
  {\bibinfo {title} {{Asymptotically flat spacetimes in three-dimensional
  higher spin gravity}},\ }\href {https://doi.org/10.1007/JHEP09(2013)016}
  {\bibfield  {journal} {\bibinfo  {journal} {JHEP}\ }\textbf {\bibinfo
  {volume} {09}},\ \bibinfo {pages} {016}},\ \Eprint
  {https://arxiv.org/abs/1307.5651} {arXiv:1307.5651 [hep-th]} \BibitemShut
  {NoStop}%
\bibitem [{\citenamefont {Ammon}\ \emph {et~al.}(2021)\citenamefont {Ammon},
  \citenamefont {Pannier},\ and\ \citenamefont {Riegler}}]{Ammon:2020fxs}%
  \BibitemOpen
  \bibfield  {author} {\bibinfo {author} {\bibfnamefont {M.}~\bibnamefont
  {Ammon}}, \bibinfo {author} {\bibfnamefont {M.}~\bibnamefont {Pannier}},\
  and\ \bibinfo {author} {\bibfnamefont {M.}~\bibnamefont {Riegler}},\
  }\bibfield  {title} {\bibinfo {title} {{Scalar Fields in 3D Asymptotically
  Flat Higher-Spin Gravity}},\ }\href
  {https://doi.org/10.1088/1751-8121/abdbc6} {\bibfield  {journal} {\bibinfo
  {journal} {J. Phys. A}\ }\textbf {\bibinfo {volume} {54}},\ \bibinfo {pages}
  {105401} (\bibinfo {year} {2021})},\ \Eprint
  {https://arxiv.org/abs/2009.14210} {arXiv:2009.14210 [hep-th]} \BibitemShut
  {NoStop}%
\bibitem [{\citenamefont {Neckam}(2023)}]{Neckam:2023}%
  \BibitemOpen
  \bibfield  {author} {\bibinfo {author} {\bibfnamefont {P.}~\bibnamefont
  {Neckam}},\ }\emph {\bibinfo {title} {Spin-3-Gravity in Three-Dimensional
  Flat Space}},\ \href {https://doi.org/10.25365/thesis.74353} {Master's
  thesis},\ \bibinfo  {school} {University of Vienna} (\bibinfo {year}
  {2023})\BibitemShut {NoStop}%
\bibitem [{Note4()}]{Note4}%
  \BibitemOpen
  \bibinfo {note} {S. Fredenhagen, K.Mkrtchyan, P. Neckam, work in
  progress.\label {FKN}}\BibitemShut {NoStop}%
\bibitem [{Note5()}]{Note5}%
  \BibitemOpen
  \bibinfo {note} {Note that in terms of the generalized Kronecker delta
  $\delta _{\beta _1\protect \dots \beta _n}^{\alpha _1\protect \dots \alpha
  _n} = n!\delta ^{\alpha _1}_{[ \beta _1}\protect \dots \delta ^{\alpha
  _n}_{\beta _n ]}$ this means $\epsilon _{\mu \nu \rho }=\delta _{\mu \nu \rho
  }^{012}$ and $\epsilon ^{\mu \nu \rho }=-\delta ^{\mu \nu \rho
  }_{012}$.}\BibitemShut {Stop}%
\bibitem [{Note6()}]{Note6}%
  \BibitemOpen
  \bibinfo {note} {In general, sending some of the coupling constants to zero
  might not be possible in the full non-linear theory, where the independent
  coupling constants may be related for the consistency of the full theory.
  However, for the classification of all nontrivial deformations of free
  action, we do not require the existence of a full nonlinear theory that uses
  the vertices thus found.}\BibitemShut {Stop}%
\bibitem [{\citenamefont {Fredenhagen}\ \emph {et~al.}(2020)\citenamefont
  {Fredenhagen}, \citenamefont {Kr\"uger},\ and\ \citenamefont
  {Mkrtchyan}}]{Fredenhagen:2019lsz}%
  \BibitemOpen
  \bibfield  {author} {\bibinfo {author} {\bibfnamefont {S.}~\bibnamefont
  {Fredenhagen}}, \bibinfo {author} {\bibfnamefont {O.}~\bibnamefont
  {Kr\"uger}},\ and\ \bibinfo {author} {\bibfnamefont {K.}~\bibnamefont
  {Mkrtchyan}},\ }\bibfield  {title} {\bibinfo {title} {{Restrictions for
  $n$-Point Vertices in Higher-Spin Theories}},\ }\href
  {https://doi.org/10.1007/JHEP06(2020)118} {\bibfield  {journal} {\bibinfo
  {journal} {JHEP}\ }\textbf {\bibinfo {volume} {06}},\ \bibinfo {pages}
  {118}},\ \Eprint {https://arxiv.org/abs/1912.13476} {arXiv:1912.13476
  [hep-th]} \BibitemShut {NoStop}%
\bibitem [{Note7()}]{Note7}%
  \BibitemOpen
  \bibinfo {note} {With a small abuse of the terminology, we use the term spin
  $s$ for the massless Fronsdal fields in three dimensions, described by
  symmetric tensors of rank $s$, and correspondingly, spin $s+1/2$, for
  Fang--Fronsdal fields, described by symmetric tensor-spinors of rank
  $s$.}\BibitemShut {Stop}%
\bibitem [{Note8()}]{Note8}%
  \BibitemOpen
  \bibinfo {note} {We use round brackets for symmetrization with weight one
  (e.g., $A_{(\mu \nu )}=\protect \frac 12 (A_{\mu \nu }+A_{\nu \mu })$), and
  square brackets for anti-symmetrization with weight one (e.g., $A_{[\mu \nu
  ]}=\protect \frac 12 (A_{\mu \nu }-A_{\nu \mu })$).}\BibitemShut {Stop}%
\bibitem [{\citenamefont {Fredenhagen}\ \emph {et~al.}()\citenamefont
  {Fredenhagen}, \citenamefont {Lausch},\ and\ \citenamefont
  {Mkrtchyan}}]{FLM2}%
  \BibitemOpen
  \bibfield  {author} {\bibinfo {author} {\bibfnamefont {S.}~\bibnamefont
  {Fredenhagen}}, \bibinfo {author} {\bibfnamefont {F.}~\bibnamefont
  {Lausch}},\ and\ \bibinfo {author} {\bibfnamefont {K.}~\bibnamefont
  {Mkrtchyan}},\ }\bibfield  {title} {\bibinfo {title} {{in preparation}},\
  }\href@noop {} {\ }\BibitemShut {NoStop}%
\bibitem [{\citenamefont {Lausch}(2022)}]{Lausch:2022}%
  \BibitemOpen
  \bibfield  {author} {\bibinfo {author} {\bibfnamefont {F.}~\bibnamefont
  {Lausch}},\ }\emph {\bibinfo {title} {Fermionic interaction Vertices in
  Higher-Spin Theory with a Focus on quartic interaction Terms}},\ \href
  {https://doi.org/10.25365/thesis.71925} {Master's thesis},\ \bibinfo
  {school} {University of Vienna} (\bibinfo {year} {2022})\BibitemShut
  {NoStop}%
\bibitem [{\citenamefont {Campoleoni}\ \emph {et~al.}(2017)\citenamefont
  {Campoleoni}, \citenamefont {Henneaux}, \citenamefont {H\"ortner},\ and\
  \citenamefont {Leonard}}]{Campoleoni:2017vds}%
  \BibitemOpen
  \bibfield  {author} {\bibinfo {author} {\bibfnamefont {A.}~\bibnamefont
  {Campoleoni}}, \bibinfo {author} {\bibfnamefont {M.}~\bibnamefont
  {Henneaux}}, \bibinfo {author} {\bibfnamefont {S.}~\bibnamefont
  {H\"ortner}},\ and\ \bibinfo {author} {\bibfnamefont {A.}~\bibnamefont
  {Leonard}},\ }\bibfield  {title} {\bibinfo {title} {{Higher-spin charges in
  Hamiltonian form. II. Fermi fields}},\ }\href
  {https://doi.org/10.1007/JHEP02(2017)058} {\bibfield  {journal} {\bibinfo
  {journal} {JHEP}\ }\textbf {\bibinfo {volume} {02}},\ \bibinfo {pages}
  {058}},\ \Eprint {https://arxiv.org/abs/1701.05526} {arXiv:1701.05526
  [hep-th]} \BibitemShut {NoStop}%
\bibitem [{\citenamefont {Achucarro}\ and\ \citenamefont
  {Townsend}(1986)}]{Achucarro:1986uwr}%
  \BibitemOpen
  \bibfield  {author} {\bibinfo {author} {\bibfnamefont {A.}~\bibnamefont
  {Achucarro}}\ and\ \bibinfo {author} {\bibfnamefont {P.~K.}\ \bibnamefont
  {Townsend}},\ }\bibfield  {title} {\bibinfo {title} {{A Chern-Simons Action
  for Three-Dimensional anti-De Sitter Supergravity Theories}},\ }\href
  {https://doi.org/10.1016/0370-2693(86)90140-1} {\bibfield  {journal}
  {\bibinfo  {journal} {Phys. Lett. B}\ }\textbf {\bibinfo {volume} {180}},\
  \bibinfo {pages} {89} (\bibinfo {year} {1986})}\BibitemShut {NoStop}%
\bibitem [{\citenamefont {Achucarro}\ and\ \citenamefont
  {Townsend}(1989)}]{Achucarro:1989gm}%
  \BibitemOpen
  \bibfield  {author} {\bibinfo {author} {\bibfnamefont {A.}~\bibnamefont
  {Achucarro}}\ and\ \bibinfo {author} {\bibfnamefont {P.~K.}\ \bibnamefont
  {Townsend}},\ }\bibfield  {title} {\bibinfo {title} {{Extended Supergravities
  in $d$ = (2+1) as {Chern-Simons} Theories}},\ }\href
  {https://doi.org/10.1016/0370-2693(89)90423-1} {\bibfield  {journal}
  {\bibinfo  {journal} {Phys. Lett. B}\ }\textbf {\bibinfo {volume} {229}},\
  \bibinfo {pages} {383} (\bibinfo {year} {1989})}\BibitemShut {NoStop}%
\bibitem [{\citenamefont {Witten}(1988)}]{Witten:1988hc}%
  \BibitemOpen
  \bibfield  {author} {\bibinfo {author} {\bibfnamefont {E.}~\bibnamefont
  {Witten}},\ }\bibfield  {title} {\bibinfo {title} {{(2+1)-Dimensional Gravity
  as an Exactly Soluble System}},\ }\href
  {https://doi.org/10.1016/0550-3213(88)90143-5} {\bibfield  {journal}
  {\bibinfo  {journal} {Nucl. Phys. B}\ }\textbf {\bibinfo {volume} {311}},\
  \bibinfo {pages} {46} (\bibinfo {year} {1988})}\BibitemShut {NoStop}%
\bibitem [{\citenamefont {Banados}\ \emph {et~al.}(1998)\citenamefont
  {Banados}, \citenamefont {Bautier}, \citenamefont {Coussaert}, \citenamefont
  {Henneaux},\ and\ \citenamefont {Ortiz}}]{Banados:1998pi}%
  \BibitemOpen
  \bibfield  {author} {\bibinfo {author} {\bibfnamefont {M.}~\bibnamefont
  {Banados}}, \bibinfo {author} {\bibfnamefont {K.}~\bibnamefont {Bautier}},
  \bibinfo {author} {\bibfnamefont {O.}~\bibnamefont {Coussaert}}, \bibinfo
  {author} {\bibfnamefont {M.}~\bibnamefont {Henneaux}},\ and\ \bibinfo
  {author} {\bibfnamefont {M.}~\bibnamefont {Ortiz}},\ }\bibfield  {title}
  {\bibinfo {title} {{Anti-de Sitter / CFT correspondence in three-dimensional
  supergravity}},\ }\href {https://doi.org/10.1103/PhysRevD.58.085020}
  {\bibfield  {journal} {\bibinfo  {journal} {Phys. Rev. D}\ }\textbf {\bibinfo
  {volume} {58}},\ \bibinfo {pages} {085020} (\bibinfo {year} {1998})},\
  \Eprint {https://arxiv.org/abs/hep-th/9805165} {arXiv:hep-th/9805165}
  \BibitemShut {NoStop}%
\bibitem [{\citenamefont {Henneaux}\ \emph {et~al.}(2000)\citenamefont
  {Henneaux}, \citenamefont {Maoz},\ and\ \citenamefont
  {Schwimmer}}]{Henneaux:1999ib}%
  \BibitemOpen
  \bibfield  {author} {\bibinfo {author} {\bibfnamefont {M.}~\bibnamefont
  {Henneaux}}, \bibinfo {author} {\bibfnamefont {L.}~\bibnamefont {Maoz}},\
  and\ \bibinfo {author} {\bibfnamefont {A.}~\bibnamefont {Schwimmer}},\
  }\bibfield  {title} {\bibinfo {title} {{Asymptotic dynamics and asymptotic
  symmetries of three-dimensional extended AdS supergravity}},\ }\href
  {https://doi.org/10.1006/aphy.2000.5994} {\bibfield  {journal} {\bibinfo
  {journal} {Annals Phys.}\ }\textbf {\bibinfo {volume} {282}},\ \bibinfo
  {pages} {31} (\bibinfo {year} {2000})},\ \Eprint
  {https://arxiv.org/abs/hep-th/9910013} {arXiv:hep-th/9910013} \BibitemShut
  {NoStop}%
\bibitem [{\citenamefont {Hyakutake}(2013)}]{Hyakutake:2012uv}%
  \BibitemOpen
  \bibfield  {author} {\bibinfo {author} {\bibfnamefont {Y.}~\bibnamefont
  {Hyakutake}},\ }\bibfield  {title} {\bibinfo {title} {{Super Virasoro Algebra
  From Supergravity}},\ }\href {https://doi.org/10.1103/PhysRevD.87.045028}
  {\bibfield  {journal} {\bibinfo  {journal} {Phys. Rev. D}\ }\textbf {\bibinfo
  {volume} {87}},\ \bibinfo {pages} {045028} (\bibinfo {year} {2013})},\
  \Eprint {https://arxiv.org/abs/1211.3547} {arXiv:1211.3547 [hep-th]}
  \BibitemShut {NoStop}%
\bibitem [{\citenamefont {Henneaux}\ \emph
  {et~al.}(2012{\natexlab{a}})\citenamefont {Henneaux}, \citenamefont
  {Lucena~G\'omez}, \citenamefont {Park},\ and\ \citenamefont
  {Rey}}]{Henneaux:2012ny}%
  \BibitemOpen
  \bibfield  {author} {\bibinfo {author} {\bibfnamefont {M.}~\bibnamefont
  {Henneaux}}, \bibinfo {author} {\bibfnamefont {G.}~\bibnamefont
  {Lucena~G\'omez}}, \bibinfo {author} {\bibfnamefont {J.}~\bibnamefont
  {Park}},\ and\ \bibinfo {author} {\bibfnamefont {S.-J.}\ \bibnamefont
  {Rey}},\ }\bibfield  {title} {\bibinfo {title} {{Super- W(infinity)
  Asymptotic Symmetry of Higher-Spin $AdS_3$ Supergravity}},\ }\href
  {https://doi.org/10.1007/JHEP06(2012)037} {\bibfield  {journal} {\bibinfo
  {journal} {JHEP}\ }\textbf {\bibinfo {volume} {06}},\ \bibinfo {pages}
  {037}},\ \Eprint {https://arxiv.org/abs/1203.5152} {arXiv:1203.5152 [hep-th]}
  \BibitemShut {NoStop}%
\bibitem [{\citenamefont {Gunaydin}\ \emph {et~al.}(1986)\citenamefont
  {Gunaydin}, \citenamefont {Sierra},\ and\ \citenamefont
  {Townsend}}]{Gunaydin:1986fe}%
  \BibitemOpen
  \bibfield  {author} {\bibinfo {author} {\bibfnamefont {M.}~\bibnamefont
  {Gunaydin}}, \bibinfo {author} {\bibfnamefont {G.}~\bibnamefont {Sierra}},\
  and\ \bibinfo {author} {\bibfnamefont {P.~K.}\ \bibnamefont {Townsend}},\
  }\bibfield  {title} {\bibinfo {title} {{The Unitary Supermultiplets of $d=3$
  Anti-de Sitter and $d=2$ Conformal Superalgebras}},\ }\href
  {https://doi.org/10.1016/0550-3213(86)90293-2} {\bibfield  {journal}
  {\bibinfo  {journal} {Nucl. Phys. B}\ }\textbf {\bibinfo {volume} {274}},\
  \bibinfo {pages} {429} (\bibinfo {year} {1986})}\BibitemShut {NoStop}%
\bibitem [{Note9()}]{Note9}%
  \BibitemOpen
  \bibinfo {note} {Note, that the holomorphic factorization is specific to
  massless fields, corresponding to a trivial representation of one of the
  simple components of the isometry (see, e.g., \cite
  {Gwak:2015vfb}).}\BibitemShut {Stop}%
\bibitem [{\citenamefont {Aragone}\ and\ \citenamefont
  {Deser}(1984)}]{Aragone:1983sz}%
  \BibitemOpen
  \bibfield  {author} {\bibinfo {author} {\bibfnamefont {C.}~\bibnamefont
  {Aragone}}\ and\ \bibinfo {author} {\bibfnamefont {S.}~\bibnamefont
  {Deser}},\ }\bibfield  {title} {\bibinfo {title} {{Hypersymmetry in $D=3$ of
  Coupled Gravity Massless Spin 5/2 System}},\ }\href
  {https://doi.org/10.1088/0264-9381/1/2/001} {\bibfield  {journal} {\bibinfo
  {journal} {Class. Quant. Grav.}\ }\textbf {\bibinfo {volume} {1}},\ \bibinfo
  {pages} {L9} (\bibinfo {year} {1984})}\BibitemShut {NoStop}%
\bibitem [{\citenamefont {Zinoviev}(2014)}]{Zinoviev:2014sza}%
  \BibitemOpen
  \bibfield  {author} {\bibinfo {author} {\bibfnamefont {Y.~M.}\ \bibnamefont
  {Zinoviev}},\ }\bibfield  {title} {\bibinfo {title} {{Hypergravity in
  AdS$_3$}},\ }\href {https://doi.org/10.1016/j.physletb.2014.10.041}
  {\bibfield  {journal} {\bibinfo  {journal} {Phys. Lett. B}\ }\textbf
  {\bibinfo {volume} {739}},\ \bibinfo {pages} {106} (\bibinfo {year}
  {2014})},\ \Eprint {https://arxiv.org/abs/1408.2912} {arXiv:1408.2912
  [hep-th]} \BibitemShut {NoStop}%
\bibitem [{\citenamefont {Rahman}(2019)}]{Rahman:2019mra}%
  \BibitemOpen
  \bibfield  {author} {\bibinfo {author} {\bibfnamefont {R.}~\bibnamefont
  {Rahman}},\ }\bibfield  {title} {\bibinfo {title} {{The Uniqueness of
  Hypergravity}},\ }\href {https://doi.org/10.1007/JHEP11(2019)115} {\bibfield
  {journal} {\bibinfo  {journal} {JHEP}\ }\textbf {\bibinfo {volume} {11}},\
  \bibinfo {pages} {115}},\ \Eprint {https://arxiv.org/abs/1905.04109}
  {arXiv:1905.04109 [hep-th]} \BibitemShut {NoStop}%
\bibitem [{\citenamefont {Manvelyan}\ \emph {et~al.}(2010)\citenamefont
  {Manvelyan}, \citenamefont {Mkrtchyan},\ and\ \citenamefont
  {R\"uhl}}]{Manvelyan:2010jr}%
  \BibitemOpen
  \bibfield  {author} {\bibinfo {author} {\bibfnamefont {R.}~\bibnamefont
  {Manvelyan}}, \bibinfo {author} {\bibfnamefont {K.}~\bibnamefont
  {Mkrtchyan}},\ and\ \bibinfo {author} {\bibfnamefont {W.}~\bibnamefont
  {R\"uhl}},\ }\bibfield  {title} {\bibinfo {title} {{General trilinear
  interaction for arbitrary even higher spin gauge fields}},\ }\href
  {https://doi.org/10.1016/j.nuclphysb.2010.04.019} {\bibfield  {journal}
  {\bibinfo  {journal} {Nucl. Phys.}\ }\textbf {\bibinfo {volume} {B836}},\
  \bibinfo {pages} {204} (\bibinfo {year} {2010})},\ \Eprint
  {https://arxiv.org/abs/1003.2877} {arXiv:1003.2877 [hep-th]} \BibitemShut
  {NoStop}%
\bibitem [{\citenamefont {Sagnotti}\ and\ \citenamefont
  {Taronna}(2011)}]{Sagnotti:2010at}%
  \BibitemOpen
  \bibfield  {author} {\bibinfo {author} {\bibfnamefont {A.}~\bibnamefont
  {Sagnotti}}\ and\ \bibinfo {author} {\bibfnamefont {M.}~\bibnamefont
  {Taronna}},\ }\bibfield  {title} {\bibinfo {title} {{String Lessons for
  Higher-Spin Interactions}},\ }\href
  {https://doi.org/10.1016/j.nuclphysb.2010.08.019} {\bibfield  {journal}
  {\bibinfo  {journal} {Nucl. Phys.}\ }\textbf {\bibinfo {volume} {B842}},\
  \bibinfo {pages} {299} (\bibinfo {year} {2011})},\ \Eprint
  {https://arxiv.org/abs/1006.5242} {arXiv:1006.5242 [hep-th]} \BibitemShut
  {NoStop}%
\bibitem [{\citenamefont {Joung}\ and\ \citenamefont
  {Taronna}(2012)}]{Joung:2011ww}%
  \BibitemOpen
  \bibfield  {author} {\bibinfo {author} {\bibfnamefont {E.}~\bibnamefont
  {Joung}}\ and\ \bibinfo {author} {\bibfnamefont {M.}~\bibnamefont
  {Taronna}},\ }\bibfield  {title} {\bibinfo {title} {{Cubic interactions of
  massless higher spins in (A)dS: metric-like approach}},\ }\href
  {https://doi.org/10.1016/j.nuclphysb.2012.03.013} {\bibfield  {journal}
  {\bibinfo  {journal} {Nucl. Phys.}\ }\textbf {\bibinfo {volume} {B861}},\
  \bibinfo {pages} {145} (\bibinfo {year} {2012})},\ \Eprint
  {https://arxiv.org/abs/1110.5918} {arXiv:1110.5918 [hep-th]} \BibitemShut
  {NoStop}%
\bibitem [{\citenamefont {Henneaux}\ \emph
  {et~al.}(2012{\natexlab{b}})\citenamefont {Henneaux}, \citenamefont
  {Lucena~G\'omez},\ and\ \citenamefont {Rahman}}]{Henneaux:2012wg}%
  \BibitemOpen
  \bibfield  {author} {\bibinfo {author} {\bibfnamefont {M.}~\bibnamefont
  {Henneaux}}, \bibinfo {author} {\bibfnamefont {G.}~\bibnamefont
  {Lucena~G\'omez}},\ and\ \bibinfo {author} {\bibfnamefont {R.}~\bibnamefont
  {Rahman}},\ }\bibfield  {title} {\bibinfo {title} {{Higher-Spin Fermionic
  Gauge Fields and Their Electromagnetic Coupling}},\ }\href
  {https://doi.org/10.1007/JHEP08(2012)093} {\bibfield  {journal} {\bibinfo
  {journal} {JHEP}\ }\textbf {\bibinfo {volume} {08}},\ \bibinfo {pages}
  {093}},\ \Eprint {https://arxiv.org/abs/1206.1048} {arXiv:1206.1048 [hep-th]}
  \BibitemShut {NoStop}%
\bibitem [{\citenamefont {Henneaux}\ \emph {et~al.}(2014)\citenamefont
  {Henneaux}, \citenamefont {Lucena~G\'omez},\ and\ \citenamefont
  {Rahman}}]{Henneaux:2013gba}%
  \BibitemOpen
  \bibfield  {author} {\bibinfo {author} {\bibfnamefont {M.}~\bibnamefont
  {Henneaux}}, \bibinfo {author} {\bibfnamefont {G.}~\bibnamefont
  {Lucena~G\'omez}},\ and\ \bibinfo {author} {\bibfnamefont {R.}~\bibnamefont
  {Rahman}},\ }\bibfield  {title} {\bibinfo {title} {{Gravitational
  Interactions of Higher-Spin Fermions}},\ }\href
  {https://doi.org/10.1007/JHEP01(2014)087} {\bibfield  {journal} {\bibinfo
  {journal} {JHEP}\ }\textbf {\bibinfo {volume} {01}},\ \bibinfo {pages}
  {087}},\ \Eprint {https://arxiv.org/abs/1310.5152} {arXiv:1310.5152 [hep-th]}
  \BibitemShut {NoStop}%
\bibitem [{\citenamefont {Conde}\ \emph {et~al.}(2016)\citenamefont {Conde},
  \citenamefont {Joung},\ and\ \citenamefont {Mkrtchyan}}]{Conde:2016izb}%
  \BibitemOpen
  \bibfield  {author} {\bibinfo {author} {\bibfnamefont {E.}~\bibnamefont
  {Conde}}, \bibinfo {author} {\bibfnamefont {E.}~\bibnamefont {Joung}},\ and\
  \bibinfo {author} {\bibfnamefont {K.}~\bibnamefont {Mkrtchyan}},\ }\bibfield
  {title} {\bibinfo {title} {{Spinor-Helicity Three-Point Amplitudes from Local
  Cubic Interactions}},\ }\href {https://doi.org/10.1007/JHEP08(2016)040}
  {\bibfield  {journal} {\bibinfo  {journal} {JHEP}\ }\textbf {\bibinfo
  {volume} {08}},\ \bibinfo {pages} {040}},\ \Eprint
  {https://arxiv.org/abs/1605.07402} {arXiv:1605.07402 [hep-th]} \BibitemShut
  {NoStop}%
\bibitem [{\citenamefont {Francia}\ \emph {et~al.}(2017)\citenamefont
  {Francia}, \citenamefont {Monaco},\ and\ \citenamefont
  {Mkrtchyan}}]{Francia:2016weg}%
  \BibitemOpen
  \bibfield  {author} {\bibinfo {author} {\bibfnamefont {D.}~\bibnamefont
  {Francia}}, \bibinfo {author} {\bibfnamefont {G.~L.}\ \bibnamefont
  {Monaco}},\ and\ \bibinfo {author} {\bibfnamefont {K.}~\bibnamefont
  {Mkrtchyan}},\ }\bibfield  {title} {\bibinfo {title} {{Cubic interactions of
  Maxwell-like higher spins}},\ }\href
  {https://doi.org/10.1007/JHEP04(2017)068} {\bibfield  {journal} {\bibinfo
  {journal} {JHEP}\ }\textbf {\bibinfo {volume} {04}},\ \bibinfo {pages}
  {068}},\ \Eprint {https://arxiv.org/abs/1611.00292} {arXiv:1611.00292
  [hep-th]} \BibitemShut {NoStop}%
\bibitem [{\citenamefont {Sleight}\ and\ \citenamefont
  {Taronna}(2018{\natexlab{b}})}]{Sleight:2017cax}%
  \BibitemOpen
  \bibfield  {author} {\bibinfo {author} {\bibfnamefont {C.}~\bibnamefont
  {Sleight}}\ and\ \bibinfo {author} {\bibfnamefont {M.}~\bibnamefont
  {Taronna}},\ }\bibfield  {title} {\bibinfo {title} {{Feynman rules for
  higher-spin gauge fields on AdS$_{d+1}$}},\ }\href
  {https://doi.org/10.1007/JHEP01(2018)060} {\bibfield  {journal} {\bibinfo
  {journal} {JHEP}\ }\textbf {\bibinfo {volume} {01}},\ \bibinfo {pages}
  {060}},\ \Eprint {https://arxiv.org/abs/1708.08668} {arXiv:1708.08668
  [hep-th]} \BibitemShut {NoStop}%
\bibitem [{\citenamefont {Joung}\ and\ \citenamefont
  {Taronna}(2020)}]{Joung:2019wbl}%
  \BibitemOpen
  \bibfield  {author} {\bibinfo {author} {\bibfnamefont {E.}~\bibnamefont
  {Joung}}\ and\ \bibinfo {author} {\bibfnamefont {M.}~\bibnamefont
  {Taronna}},\ }\bibfield  {title} {\bibinfo {title} {{A note on higher-order
  vertices of higher-spin fields in flat and (A)dS space}},\ }\href
  {https://doi.org/10.1007/JHEP09(2020)171} {\bibfield  {journal} {\bibinfo
  {journal} {JHEP}\ }\textbf {\bibinfo {volume} {09}},\ \bibinfo {pages}
  {171}},\ \Eprint {https://arxiv.org/abs/1912.12357} {arXiv:1912.12357
  [hep-th]} \BibitemShut {NoStop}%
\bibitem [{\citenamefont {Bengtsson}\ \emph {et~al.}(1983)\citenamefont
  {Bengtsson}, \citenamefont {Bengtsson},\ and\ \citenamefont
  {Brink}}]{Bengtsson:1983pd}%
  \BibitemOpen
  \bibfield  {author} {\bibinfo {author} {\bibfnamefont {A.~K.~H.}\
  \bibnamefont {Bengtsson}}, \bibinfo {author} {\bibfnamefont {I.}~\bibnamefont
  {Bengtsson}},\ and\ \bibinfo {author} {\bibfnamefont {L.}~\bibnamefont
  {Brink}},\ }\bibfield  {title} {\bibinfo {title} {{Cubic Interaction Terms
  for Arbitrary Spin}},\ }\href {https://doi.org/10.1016/0550-3213(83)90140-2}
  {\bibfield  {journal} {\bibinfo  {journal} {Nucl. Phys.}\ }\textbf {\bibinfo
  {volume} {B227}},\ \bibinfo {pages} {31} (\bibinfo {year}
  {1983})}\BibitemShut {NoStop}%
\bibitem [{\citenamefont {Bengtsson}\ \emph {et~al.}(1987)\citenamefont
  {Bengtsson}, \citenamefont {Bengtsson},\ and\ \citenamefont
  {Linden}}]{Bengtsson:1986kh}%
  \BibitemOpen
  \bibfield  {author} {\bibinfo {author} {\bibfnamefont {A.~K.~H.}\
  \bibnamefont {Bengtsson}}, \bibinfo {author} {\bibfnamefont {I.}~\bibnamefont
  {Bengtsson}},\ and\ \bibinfo {author} {\bibfnamefont {N.}~\bibnamefont
  {Linden}},\ }\bibfield  {title} {\bibinfo {title} {{Interacting Higher Spin
  Gauge Fields on the Light Front}},\ }\href
  {https://doi.org/10.1088/0264-9381/4/5/028} {\bibfield  {journal} {\bibinfo
  {journal} {Class. Quant. Grav.}\ }\textbf {\bibinfo {volume} {4}},\ \bibinfo
  {pages} {1333} (\bibinfo {year} {1987})}\BibitemShut {NoStop}%
\bibitem [{\citenamefont {Fradkin}\ and\ \citenamefont
  {Metsaev}(1991)}]{Fradkin:1991iy}%
  \BibitemOpen
  \bibfield  {author} {\bibinfo {author} {\bibfnamefont {E.~S.}\ \bibnamefont
  {Fradkin}}\ and\ \bibinfo {author} {\bibfnamefont {R.~R.}\ \bibnamefont
  {Metsaev}},\ }\bibfield  {title} {\bibinfo {title} {{A Cubic interaction of
  totally symmetric massless representations of the Lorentz group in arbitrary
  dimensions}},\ }\href {https://doi.org/10.1088/0264-9381/8/4/004} {\bibfield
  {journal} {\bibinfo  {journal} {Class. Quant. Grav.}\ }\textbf {\bibinfo
  {volume} {8}},\ \bibinfo {pages} {L89} (\bibinfo {year} {1991})}\BibitemShut
  {NoStop}%
\bibitem [{\citenamefont {Metsaev}(2006)}]{Metsaev:2005ar}%
  \BibitemOpen
  \bibfield  {author} {\bibinfo {author} {\bibfnamefont {R.~R.}\ \bibnamefont
  {Metsaev}},\ }\bibfield  {title} {\bibinfo {title} {{Cubic interaction
  vertices of massive and massless higher spin fields}},\ }\href
  {https://doi.org/10.1016/j.nuclphysb.2006.10.002} {\bibfield  {journal}
  {\bibinfo  {journal} {Nucl. Phys.}\ }\textbf {\bibinfo {volume} {B759}},\
  \bibinfo {pages} {147} (\bibinfo {year} {2006})},\ \Eprint
  {https://arxiv.org/abs/hep-th/0512342} {arXiv:hep-th/0512342 [hep-th]}
  \BibitemShut {NoStop}%
\bibitem [{\citenamefont {Metsaev}(2012)}]{Metsaev:2007rn}%
  \BibitemOpen
  \bibfield  {author} {\bibinfo {author} {\bibfnamefont {R.~R.}\ \bibnamefont
  {Metsaev}},\ }\bibfield  {title} {\bibinfo {title} {{Cubic interaction
  vertices for fermionic and bosonic arbitrary spin fields}},\ }\href
  {https://doi.org/10.1016/j.nuclphysb.2012.01.022} {\bibfield  {journal}
  {\bibinfo  {journal} {Nucl. Phys.}\ }\textbf {\bibinfo {volume} {B859}},\
  \bibinfo {pages} {13} (\bibinfo {year} {2012})},\ \Eprint
  {https://arxiv.org/abs/0712.3526} {arXiv:0712.3526 [hep-th]} \BibitemShut
  {NoStop}%
\bibitem [{\citenamefont {Fradkin}\ and\ \citenamefont
  {Vasiliev}(1987{\natexlab{b}})}]{Fradkin:1986qy}%
  \BibitemOpen
  \bibfield  {author} {\bibinfo {author} {\bibfnamefont {E.~S.}\ \bibnamefont
  {Fradkin}}\ and\ \bibinfo {author} {\bibfnamefont {M.~A.}\ \bibnamefont
  {Vasiliev}},\ }\bibfield  {title} {\bibinfo {title} {{Cubic Interaction in
  Extended Theories of Massless Higher Spin Fields}},\ }\href
  {https://doi.org/10.1016/0550-3213(87)90469-X} {\bibfield  {journal}
  {\bibinfo  {journal} {Nucl. Phys.}\ }\textbf {\bibinfo {volume} {B291}},\
  \bibinfo {pages} {141} (\bibinfo {year} {1987}{\natexlab{b}})}\BibitemShut
  {NoStop}%
\bibitem [{\citenamefont {Vasiliev}(2012)}]{Vasiliev:2011knf}%
  \BibitemOpen
  \bibfield  {author} {\bibinfo {author} {\bibfnamefont {M.~A.}\ \bibnamefont
  {Vasiliev}},\ }\bibfield  {title} {\bibinfo {title} {{Cubic Vertices for
  Symmetric Higher-Spin Gauge Fields in $(A)dS_d$}},\ }\href
  {https://doi.org/10.1016/j.nuclphysb.2012.04.012} {\bibfield  {journal}
  {\bibinfo  {journal} {Nucl. Phys. B}\ }\textbf {\bibinfo {volume} {862}},\
  \bibinfo {pages} {341} (\bibinfo {year} {2012})},\ \Eprint
  {https://arxiv.org/abs/1108.5921} {arXiv:1108.5921 [hep-th]} \BibitemShut
  {NoStop}%
\bibitem [{\citenamefont {Boulanger}\ \emph {et~al.}(2013)\citenamefont
  {Boulanger}, \citenamefont {Ponomarev},\ and\ \citenamefont
  {Skvortsov}}]{Boulanger:2012dx}%
  \BibitemOpen
  \bibfield  {author} {\bibinfo {author} {\bibfnamefont {N.}~\bibnamefont
  {Boulanger}}, \bibinfo {author} {\bibfnamefont {D.}~\bibnamefont
  {Ponomarev}},\ and\ \bibinfo {author} {\bibfnamefont {E.~D.}\ \bibnamefont
  {Skvortsov}},\ }\bibfield  {title} {\bibinfo {title} {{Non-abelian cubic
  vertices for higher-spin fields in anti-de Sitter space}},\ }\href
  {https://doi.org/10.1007/JHEP05(2013)008} {\bibfield  {journal} {\bibinfo
  {journal} {JHEP}\ }\textbf {\bibinfo {volume} {05}},\ \bibinfo {pages}
  {008}},\ \Eprint {https://arxiv.org/abs/1211.6979} {arXiv:1211.6979 [hep-th]}
  \BibitemShut {NoStop}%
\bibitem [{\citenamefont {Khabarov}\ and\ \citenamefont
  {Zinoviev}(2020)}]{Khabarov:2020bgr}%
  \BibitemOpen
  \bibfield  {author} {\bibinfo {author} {\bibfnamefont {M.~V.}\ \bibnamefont
  {Khabarov}}\ and\ \bibinfo {author} {\bibfnamefont {Y.~M.}\ \bibnamefont
  {Zinoviev}},\ }\bibfield  {title} {\bibinfo {title} {{Massless higher spin
  cubic vertices in flat four dimensional space}},\ }\href
  {https://doi.org/10.1007/JHEP08(2020)112} {\bibfield  {journal} {\bibinfo
  {journal} {JHEP}\ }\textbf {\bibinfo {volume} {08}},\ \bibinfo {pages}
  {112}},\ \Eprint {https://arxiv.org/abs/2005.09851} {arXiv:2005.09851
  [hep-th]} \BibitemShut {NoStop}%
\bibitem [{\citenamefont {Sleight}(2017)}]{Sleight:2016hyl}%
  \BibitemOpen
  \bibfield  {author} {\bibinfo {author} {\bibfnamefont {C.}~\bibnamefont
  {Sleight}},\ }\bibfield  {title} {\bibinfo {title} {{Interactions in
  Higher-Spin Gravity: a Holographic Perspective}},\ }\href
  {https://doi.org/10.1088/1751-8121/aa820c} {\bibfield  {journal} {\bibinfo
  {journal} {J. Phys. A}\ }\textbf {\bibinfo {volume} {50}},\ \bibinfo {pages}
  {383001} (\bibinfo {year} {2017})},\ \Eprint
  {https://arxiv.org/abs/1610.01318} {arXiv:1610.01318 [hep-th]} \BibitemShut
  {NoStop}%
\bibitem [{\citenamefont {Ponomarev}(2023)}]{Ponomarev:2022vjb}%
  \BibitemOpen
  \bibfield  {author} {\bibinfo {author} {\bibfnamefont {D.}~\bibnamefont
  {Ponomarev}},\ }\bibfield  {title} {\bibinfo {title} {{Basic Introduction to
  Higher-Spin Theories}},\ }\href {https://doi.org/10.1007/s10773-023-05399-5}
  {\bibfield  {journal} {\bibinfo  {journal} {Int. J. Theor. Phys.}\ }\textbf
  {\bibinfo {volume} {62}},\ \bibinfo {pages} {146} (\bibinfo {year} {2023})},\
  \Eprint {https://arxiv.org/abs/2206.15385} {arXiv:2206.15385 [hep-th]}
  \BibitemShut {NoStop}%
\bibitem [{\citenamefont {Bengtsson}(2023)}]{Bengtsson:2023ucs}%
  \BibitemOpen
  \bibfield  {author} {\bibinfo {author} {\bibfnamefont {A.}~\bibnamefont
  {Bengtsson}},\ }\href {https://doi.org/10.1515/9783110675528} {\emph
  {\bibinfo {title} {{Interactions}}}},\ Higher Spin Field Theory\ (\bibinfo
  {publisher} {De Gruyter},\ \bibinfo {year} {2023})\BibitemShut {NoStop}%
\bibitem [{\citenamefont {Pekar}(2023)}]{Pekar:2023nev}%
  \BibitemOpen
  \bibfield  {author} {\bibinfo {author} {\bibfnamefont {S.}~\bibnamefont
  {Pekar}},\ }\bibfield  {title} {\bibinfo {title} {{Introduction to
  higher-spin theories}},\ }\href {https://doi.org/10.22323/1.435.0004}
  {\bibfield  {journal} {\bibinfo  {journal} {PoS}\ }\textbf {\bibinfo {volume}
  {Modave2022}},\ \bibinfo {pages} {004} (\bibinfo {year} {2023})}\BibitemShut
  {NoStop}%
\bibitem [{\citenamefont {Campoleoni}\ and\ \citenamefont
  {Fredenhagen}(2024)}]{Campoleoni:2024ced}%
  \BibitemOpen
  \bibfield  {author} {\bibinfo {author} {\bibfnamefont {A.}~\bibnamefont
  {Campoleoni}}\ and\ \bibinfo {author} {\bibfnamefont {S.}~\bibnamefont
  {Fredenhagen}},\ }\bibfield  {title} {\bibinfo {title} {{Higher-spin gauge
  theories in three spacetime dimensions}}\ }(\bibinfo {year} {2024})\ \Eprint
  {https://arxiv.org/abs/2403.16567} {arXiv:2403.16567 [hep-th]} \BibitemShut
  {NoStop}%
\bibitem [{\citenamefont {Bonezzi}\ \emph {et~al.}(2016)\citenamefont
  {Bonezzi}, \citenamefont {Boulanger}, \citenamefont {Sezgin},\ and\
  \citenamefont {Sundell}}]{Bonezzi:2015igv}%
  \BibitemOpen
  \bibfield  {author} {\bibinfo {author} {\bibfnamefont {R.}~\bibnamefont
  {Bonezzi}}, \bibinfo {author} {\bibfnamefont {N.}~\bibnamefont {Boulanger}},
  \bibinfo {author} {\bibfnamefont {E.}~\bibnamefont {Sezgin}},\ and\ \bibinfo
  {author} {\bibfnamefont {P.}~\bibnamefont {Sundell}},\ }\bibfield  {title}
  {\bibinfo {title} {{An Action for Matter Coupled Higher Spin Gravity in Three
  Dimensions}},\ }\href {https://doi.org/10.1007/JHEP05(2016)003} {\bibfield
  {journal} {\bibinfo  {journal} {JHEP}\ }\textbf {\bibinfo {volume} {05}},\
  \bibinfo {pages} {003}},\ \Eprint {https://arxiv.org/abs/1512.02209}
  {arXiv:1512.02209 [hep-th]} \BibitemShut {NoStop}%
\bibitem [{\citenamefont {Gwak}\ \emph {et~al.}(2016)\citenamefont {Gwak},
  \citenamefont {Joung}, \citenamefont {Mkrtchyan},\ and\ \citenamefont
  {Rey}}]{Gwak:2015vfb}%
  \BibitemOpen
  \bibfield  {author} {\bibinfo {author} {\bibfnamefont {S.}~\bibnamefont
  {Gwak}}, \bibinfo {author} {\bibfnamefont {E.}~\bibnamefont {Joung}},
  \bibinfo {author} {\bibfnamefont {K.}~\bibnamefont {Mkrtchyan}},\ and\
  \bibinfo {author} {\bibfnamefont {S.-J.}\ \bibnamefont {Rey}},\ }\bibfield
  {title} {\bibinfo {title} {{Rainbow Valley of Colored (Anti) de Sitter
  Gravity in Three Dimensions}},\ }\href
  {https://doi.org/10.1007/JHEP04(2016)055} {\bibfield  {journal} {\bibinfo
  {journal} {JHEP}\ }\textbf {\bibinfo {volume} {04}},\ \bibinfo {pages}
  {055}},\ \Eprint {https://arxiv.org/abs/1511.05220} {arXiv:1511.05220
  [hep-th]} \BibitemShut {NoStop}%
\end{thebibliography}%

\end{document}